\documentclass[10pt,aps,prx,twocolumn,notitlepage,showpacs,superscriptaddress]{revtex4-1}

\usepackage{amsthm,amsmath,amsfonts,amssymb}
\usepackage{graphicx,color}
\usepackage{bm,times}
\usepackage{epsfig,slashed,lipsum}
\usepackage{verbatim,textcomp}
\usepackage[colorlinks=true,citecolor=blue,linkcolor=blue,urlcolor=blue]{hyperref}
\numberwithin{equation}{section}

\begin{document}

\newcommand{\no}{\nonumber}
\newcommand{\p}{\partial}
\newcommand{\cl}{\mathrm{cl}}
\newcommand{\mb}{\bm}
\newcommand{\A}{\text{a}}
\newcommand{\B}{\text{b}}
\newcommand{\C}{\text{c}}
\newcommand{\D}{\text{d}}
\newcommand{\DD}{\text{D}}
\newcommand{\E}{\text{e}}
\newcommand{\BO}{\text{B}}
\newcommand{\BT}{\text{BT}}
\newcommand{\HG}{\text{HG}}
\newcommand{\UV}{\text{UV}}
\newcommand{\OO}{\text{O}}
\newcommand{\PP}{\text{P}}
\newcommand{\Pade}{\text{Pad$\acute{\text{e}}$}}
\newcommand{\SU}{\text{SU}}
\newcommand{\SO}{\text{SO}}
\newcommand{\VBS}{\text{VBS}}

\title{Deconfined criticality in the QED$_{3}$ Gross-Neveu-Yukawa model: The $1/N$ expansion revisited}

\author{Rufus Boyack}
\affiliation{Department of Physics, University of Alberta, Edmonton, Alberta T6G 2E1, Canada}
\affiliation{Theoretical Physics Institute, University of Alberta, Edmonton, Alberta T6G 2E1, Canada}

\author{Ahmed Rayyan}
\affiliation{Department of Physics, University of Alberta, Edmonton, Alberta T6G 2E1, Canada}

\author{Joseph Maciejko}
\affiliation{Department of Physics, University of Alberta, Edmonton, Alberta T6G 2E1, Canada}
\affiliation{Theoretical Physics Institute, University of Alberta, Edmonton, Alberta T6G 2E1, Canada}
\affiliation{Canadian Institute for Advanced Research, Toronto, Ontario M5G 1Z8, Canada}

\date\today

\begin{abstract}
The critical properties of the QED$_{3}$ Gross-Neveu-Yukawa (GNY) model in 2+1 dimensions with $N$ flavors of two-component Dirac fermions are computed to first order in the $1/N$ expansion. For the specific case of $N=2$, the critical point is conjectured to be dual to the N\'eel-to-valence-bond-solid (VBS) deconfined critical point of quantum antiferromagnets on the square lattice. It is found that Aslamazov-Larkin diagrams, missed by previous $\epsilon$- and $1/N$-expansion studies with four-component fermions, give important contributions to the scaling dimensions of various operators. With the inclusion of these diagrams, the resummed scaling dimensions of the adjoint fermion bilinear and scalar field at the QED$_{3}$ GNY critical point are in reasonable agreement with numerical studies of the N\'eel-to-VBS transition, in support of the duality conjecture. 
\end{abstract}

\maketitle

\section{Introduction}

Quantum field theories in 2+1 dimensions can exhibit rich phenomena that lie at the interface of condensed matter physics and high-energy physics. Three-dimensional (3D) quantum electrodynamics (QED$_{3}$) is one such theory that, while initially studied in high-energy physics as a model of dynamical chiral symmetry breaking~\cite{Appelquist_1988,Nash_1989}, 
was soon found to arise in various condensed matter contexts including high-temperature superconductivity~\cite{Affleck_1988,kim1999,Franz_2001,Franz_2002,Herbut_2002}, spin liquids~\cite{hastings2000,Rantner_Wen_2001,Rantner_Wen_2002,hermele2004,Hermele_Senthil_Fisher_2005,ran2007,
hermele2008}, and the fractional quantum Hall effect~\cite{WeiChen_1993,Vishwanath_He_2018}. Another quantum field theory that has engendered a lot of recent interest in the condensed matter community is the Gross-Neveu~\cite{Gross_Neveu_1974} (or related Gross-Neveu-Yukawa (GNY)~\cite{ZinnJustin_1991}) model, which describes the critical properties of the Mott transition in graphene~\cite{herbut2009,herbut2009b}. At the GNY critical point, bosonic order parameter fluctuations are strongly Yukawa-coupled to gapless Dirac fermions, and the critical properties of this ``relativistic'' Mott transition are not adequately captured by a purely bosonic Landau-Ginzburg-Wilson theory~\cite{QPT}.

More generally, two interesting classes of (2+1)D quantum phase transitions in condensed matter are those described by GNY-type theories with coupled bosonic and fermionic matter fields, and those described by quantum field theories in which the matter sector is supplemented by a dynamical gauge field, as in the $\mathbb{C}$P$^1$~\cite{dadda1978,wu1984} and QED$_{3}$ GNY~\cite{GraceyQED3GNY_1992} models. In this latter class a dynamical gauge field emerges because the matter fields are coarse-grained versions of fractionalized degrees of freedom. Two interesting subclasses can be further delineated based on whether the gauge field is deconfined $(i)$ in (at least) one of the two phases separated by the transition, or $(ii)$ only at the critical point itself.
The latter scenario, dubbed deconfined quantum criticality~\cite{Senthil_2004,senthil2004b}, is of particular importance as a generic mechanism allowing continuous quantum phase transitions outside the traditional Landau-Ginzburg-Wilson paradigm. Remarkably, it has recently been conjectured that certain critical points belonging to these two seemingly disparate subclasses are related by infrared dualities~\cite{Wang_2017} similar in spirit to the particle-vortex duality relating the critical points of the (2+1)D XY and Abelian Higgs models~\cite{peskin1978,dasgupta1981}. In contrast with the latter, however, the dualities in Ref.~\cite{Wang_2017} involve fermions in an essential way and can be seen as belonging to a web of new dualities with relevance to both condensed matter and high-energy physics; for a recent review, see Ref.~\cite{Senthil_2018}.

One notable instance of duality between two different deconfined critical points is a conjecture~\cite{Wang_2017} relating the N\'eel-to-valence-bond-solid (VBS) transition in quantum antiferromagnets on the 2D square lattice~\cite{Senthil_2004,senthil2004b}, 
which is in subclass $(ii)$, to the critical point of the QED$_{3}$ GNY model with two flavors of two-component Dirac fermions, which is in subclass $(i)$.
If true, this duality would imply an emergent $\SO(5)$ symmetry in both models at the critical point, as well as the following equality: 
\begin{equation}
\label{eq:Eta1}
\eta_{\text{N\'eel}}=\eta_{\text{VBS}}\protect\overset{?}{=}\eta_{\phi},
\end{equation}
where $\eta_{\text{N\'eel}}$ and $\eta_{\text{VBS}}$ denote the anomalous dimensions of the N\'eel and VBS order parameters, respectively, and $\eta_\phi$ is the scalar field anomalous dimension in QED$_{3}$ GNY. Numerical studies of the N\'eel-to-VBS transition have indeed observed the emergent $\SO(5)$ symmetry~\cite{Nahum_2015} and found values of $\eta_{\text{N\'eel}}$ and $\eta_{\VBS}$ in close agreement with one another~\cite{Sandvik_2007,Melko_2008,Nahum_2015}. An additional consequence of this duality is a relation between the inverse correlation length exponent at the N\'eel-to-VBS transition and the scaling dimension of the adjoint fermion bilinear in QED$_{3}$ GNY:
\begin{equation}
\label{eq:And1}
\nu^{-1}_\text{N\'eel-VBS}\protect\overset{?}{=}3-\Delta_{\overline{\psi}T_{A}\psi}.
\end{equation}
Finally, infrared duality between critical points requires the scaling dimension of the symmetric relevant operator tuning the transition to be the same on both sides, which further implies an equality between the (inverse) correlation length exponents at the N\'eel-to-VBS and QED$_3$ GNY critical points.

Various theoretical studies of the QED$_{3}$ GNY model have recently been undertaken to test the above conjectured equalities. A one-loop renormalization group study of the critical QED$_{3}$ GNY model in the $4-\epsilon$ expansion was first performed in Ref.~\cite{Janssen_He_2017}, followed soon by analyses at three-loop~\cite{Ihrig_Janssen_2018} and four-loop~\cite{Zerf_Marquard_2018} orders. In the latter two analyses, resummation methods were employed in order to extrapolate to the physical case of $d=3$ spacetime dimensions. While for certain loop orders and choices of resummation technique the results for some of the resummed exponents were in reasonable agreement with the duality conjecture, no clear convergence was found with increasing loop orders or across resummation methods. In particular, resummed values for the scalar field anomalous dimension $\eta_\phi$ were found to vary by as much as two orders of magnitude for different choices of resummation method at four-loop order~\cite{Zerf_Marquard_2018}, rendering a meaningful test of Eq.~(\ref{eq:Eta1}) virtually impossible. Besides creating these difficulties of practice, the $4-\epsilon$ expansion also poses difficulties of principle due to inherent differences between spinor structures in $d=3$, the dimension one wishes to study, and $d=4$, the dimension about which one expands~\cite{dipietro2016,pietro2017}. As will be explained in detail below (see Sec.~\ref{sec:diracalgebra}), the theory studied in the $4-\epsilon$ expansion~\cite{Janssen_He_2017,Ihrig_Janssen_2018,Zerf_Marquard_2018} and extrapolated to $d=3$ is, strictly speaking, {\it not} the same as the theory appearing in the duality conjecture~\cite{Wang_2017}. These issues can in principle be avoided by performing an expansion in $1/N$ where $N$ is the number of flavors of Dirac fermions. The QED$_{3}$ GNY model has been studied this way previously at $\mathcal{O}(1/N)$~\cite{GraceyQED3GNY_1992, AlanneBlasiQED3GNY_2018} and $\mathcal{O}(1/N^2)$~\cite{GraceyQED3GNY_2018}, but in (fixed) continuous $d$ which necessitates the use of four-component Dirac spinors. Once again this corresponds to a different theory than that appearing in the duality conjecture, even when setting $d=3$.

Motivated by this current state of affairs, in this paper we revisit the $1/N$ expansion of the QED$_3$ GNY model in fixed $d=3$, defined as the model appearing in the duality conjecture of Ref.~\cite{Wang_2017}, and find additional (Aslamazov-Larkin) diagrams arising from Lorentz tensor structures peculiar to the use of two-component spinors in $d=3$ [see Eq.~(\ref{eq:Gamma3})].
For concreteness, we calculate the adjoint $\Delta_{\overline{\psi}T_A\psi}$ and singlet $\Delta_{\overline{\psi}\psi}$ fermion bilinear scaling dimensions, the scalar anomalous dimension $\eta_\phi$, and the inverse correlation length exponent $\nu^{-1}$, all to $\mathcal{O}(1/N)$ in fixed $d=3$ with $N$ flavors of two-component Dirac spinors. We apply Pad\'e and Borel-Pad\'e resummation methods to those results for the case where $N=2$, which applies to the duality conjecture. We find that including the Aslamazov-Larkin (AL) diagrams, our results for $\eta_\phi$ are in reasonable agreement with numerical determinations of the N\'eel and VBS order parameter anomalous dimensions at the N\'eel-to-VBS critical point, in support of the first conjectured equality, Eq.~(\ref{eq:Eta1}). However, it is not possible to resum $\nu^{-1}$ in the QED$_3$ GNY theory at this order. 
Nonetheless, the additional contributions that we find to the latter quantity, which now render it positive, are important facets of the problem that hitherto have been omitted. Our analysis also demonstrates that at $\mathcal{O}(1/N)$ there are no new contributions to the adjoint fermion bilinear scaling dimension; our results for this quantity agree at this order with those of Ref.~\cite{GraceyQED3GNY_2018}, and upon resummation are in reasonable agreeement with the second conjectured equality, Eq.~(\ref{eq:And1}).

The structure of the paper is as follows. In Sec.~\ref{sec:Formalism} we introduce the theoretical model under investigation, define the scaling dimensions of interest, and present a summary of our results.
Following this, Sec.~\ref{sec:Fermion_CorrFunc} and Sec.~\ref{sec:Boson_CorrFunc} present the detailed derivations of these results.
Finally, in Sec.~\ref{sec:Conclusion} a brief conclusion is presented.

\section{Theoretical formalism}
\label{sec:Formalism}
\subsection{Model}

As formulated in the duality conjecture~\cite{Wang_2017}, the chiral-Ising QED$_{3}$ GNY model is a (2+1)D quantum field theory described by the Euclidean-spacetime Lagrangian:
\begin{align}
\label{eq:QED3GNY_Lagrangian}
\mathcal{L}&=\sum_{i=1}^{N}\left[\overline{\psi}_{i}\gamma_{\mu}\left(\partial_{\mu}+\frac{1}{\sqrt{N}}iA_{\mu}\right)\psi_{i}+\frac{1}{\sqrt{N}}\phi\overline{\psi}_{i}\psi_{i}\right]\no\\
&\quad+\frac{1}{4e^{2}N}F_{\mu\nu}^{2}+\frac{1}{2g^2}\phi\left(r-\partial^{2}\right)\phi+\lambda^{2}\phi^{4},
\end{align}
where the $\psi_i$, $i=1,\ldots,N$, represent $N$ flavors of {\it two-component} Dirac fermion fields ($N=2$ in the duality conjecture), and $\overline{\psi}_{i}=\psi^{\dagger}_{i}\gamma_{0}$ are the Dirac conjugate fields.
The three gamma matrices $\gamma_{\mu}$ are chosen to form a two-dimensional representation of the Clifford algebra:
 $\left\{\gamma_{\mu},\gamma_{\nu}\right\}=2\delta_{\mu\nu}\mathbb{I}_{2}$, with $\mathbb{I}_{2}$ the $2\times2$ identity matrix in spinor space and $\mu,\nu\in\{0,1,2\}$ spacetime indices.  
One particular representation is given by the three Pauli matrices. 
The real scalar field $\phi$ is coupled to the $N$ fermions via a Yukawa interaction with unit strength; 
note that the coupling constant $g$ has been absorbed into the scalar kinetic term for later convenience.
The gauge field $A_{\mu}$ is coupled to only the fermions, through the covariant derivative term, 
and again note that the coupling constant $e$ has been absorbed into the gauge-field kinetic term. 
The field-strength tensor is $F_{\mu\nu}=\p_{\mu}A_{\nu}-\p_{\nu}A_{\mu}$.
A gauge-fixing term $\mathcal{L}_{\mathrm{gf}}=\frac{1}{2e^2N\xi}\left(\partial_{\mu}A_{\mu}\right)^2$ is also added to the Lagrangian,
and for the purposes of this paper we work in the Landau gauge: $\xi=0$. 
The Yukawa interaction term above preserves the global $\SU(N)$ symmetry present in the pure-QED$_{3}$ action; as will be explained in Sec.~\ref{sec:diracalgebra}, this is an important feature of the theory that distinguishes it from the theories previously studied in the $\epsilon$-expansion~\cite{Janssen_He_2017,Ihrig_Janssen_2018,Zerf_Marquard_2018} and the $1/N$-expansion~\cite{GraceyQED3GNY_1992,GraceyQED3GNY_2018}.

The parameter $r$ is related to the square of the mass of the scalar field, with the phase transition occurring at the critical point $r=0$. 
When the scalar field acquires a nonzero vacuum expectation value the fermions will acquire a time-reversal symmetry (TRS) breaking mass term, which corresponds to a transition from a gapless algebraic spin liquid to a gapped chiral spin liquid~\cite{Janssen_He_2017}. 
There is no Chern-Simons (CS) term in the Lagrangian, since a CS term explicitly breaks TRS. In the particular model under study, it is the acquisition of a nonzero vacuum expectation value by the scalar field which leads to spontaneous TRS breaking. At the critical point itself, which is the focus of our study, TRS remains unbroken and thus a CS term cannot be generated under renormalization.

The Feynman rules can be determined from the Lagrangian in Eq.~(\ref{eq:QED3GNY_Lagrangian}); 
for the bare fermion, scalar, and gauge propagators, respectively, we obtain
\begin{align}
\left(G_{\psi}^{0}\right)_{ij}\left(q\right)&=-i\frac{\slashed{q}}{q^{2}}\delta_{ij},\label{eq:Fermion_prop}\\
G_{\phi}^{0}\left(q\right)&=\frac{g^2}{q^{2}},\label{eq:Scalar_prop}\\
\Pi_{\mu\nu}^{0}\left(q\right)&=\frac{Ne^{2}}{q^{2}}\left(\delta_{\mu\nu}-\frac{q_{\mu}q_{\nu}}{q^{2}}(1-\xi)\right).\label{eq:Gauge_prop}
\end{align}
Here, Feynman's slashed notation $\slashed{X}\equiv\gamma_{\mu}X_{\mu}$ has been utilized. Note that the scalar propagator is given for the massless case, which corresponds to $r=0$ in Eq.~(\ref{eq:QED3GNY_Lagrangian}). Similarly, the bare Yukawa and QED$_{3}$ vertices are
\begin{align}
\label{eq:Scalar_vertex}\gamma_{\phi\overline{\psi}\psi}&=\frac{1}{\sqrt{N}},\\
\label{eq:Gauge_vertex}\left(\gamma_{A\overline{\psi}\psi}\right)_\mu&=\frac{i}{\sqrt{N}}\gamma_{\mu}.
\end{align}

\subsection{Dimensional continuation and flavor symmetry}
\label{sec:diracalgebra}

As already emphasized, the model given in Eq.~(\ref{eq:QED3GNY_Lagrangian}) and appearing in the duality conjecture~\cite{Wang_2017} is defined in terms of two-component Dirac spinors $\psi_i$, with $2\times 2$ gamma matrices $\gamma_\mu$ forming an irreducible representation of the 3D Dirac algebra. Ignoring anomaly considerations, the number $N$ of flavors in Eq.~(\ref{eq:QED3GNY_Lagrangian}) can be even or odd, and the Yukawa coupling preserves the full $\SU(N)$ flavor symmetry of the pure-QED$_3$ Lagrangian. By contrast, when theories of the type described by Eq.~(\ref{eq:QED3GNY_Lagrangian}) are studied in the $d=4-\epsilon$ expansion~\cite{Janssen_He_2017,Ihrig_Janssen_2018,Zerf_Marquard_2018}, one works with $N_f$ four-component Dirac spinors $\Psi_i$ and $4\times 4$ gamma matrices $\widetilde{\gamma}_\mu$ (which form a reducible representation of the Dirac algebra in $d=3$), and the Yukawa term is chosen to be of the form $\propto\phi\sum_{i=1}^{N_{f}}\overline{\Psi}_{i}\Psi_{i}$ to preserve Lorentz invariance in four dimensions, where $\overline{\Psi}_i\equiv\Psi^{\dagger}_i\widetilde{\gamma}_0$ is the 4D Dirac conjugate. To match the field content of Eq.~(\ref{eq:QED3GNY_Lagrangian}) one requires $2N_{f}=N$, which already implies that the two theories are different as only even values of $N$ can be studied in this way.

To be more explicit, the Lagrangian in the 4D theory is thus
\begin{equation}
\mathcal{L}_{4\DD}=\sum_{i=1}^{N_{f}}\biggl[\overline{\Psi}_{i}\widetilde{\gamma}_{\mu}\left(\partial_{\mu}+ieA_{\mu}
\right)\Psi_{i}+g\phi\overline{\Psi}_{i}\Psi_{i}\biggr]+\dots.
\end{equation}
In this subsection alone we use a different choice of normalization of the gauge and Yukawa couplings. To match the Lagrangian in Eq.~(\ref{eq:QED3GNY_Lagrangian}), one must express the $N_{f}$ four-component Dirac spinors in terms of $2N_f=N$ two-component spinors as
\begin{align}
\Psi_{i}=\left(\begin{array}{c}
\psi_{i} \\
\psi_{i+N_f}
\end{array}\right),\hspace{5mm}i=1,\ldots,N_f.
\end{align} 
When reducing from four to three dimensions and expressing the theory in terms of two-component spinors, the only way to obtain a Lorentz-invariant and flavor-diagonal Yukawa coupling as in Eq.~(\ref{eq:QED3GNY_Lagrangian}) is to choose the following 4D gamma matrix representation~\cite{Kubota_2001}:
\begin{align}
\widetilde{\gamma}_{\mu}=\left(\begin{array}{cc}
\gamma_{\mu} & 0 \\
0 & -\gamma_{\mu}
\end{array}\right),\mu=0,1,2, \hspace{2mm}
\widetilde{\gamma}_{3}=\left(\begin{array}{cc}
0 & -i \\
i & 0
\end{array}\right).
\end{align} 
However, when reducing $\mathcal{L}_{4\DD}$ from four to three dimensions this gives the Lagrangian
\begin{align}
\label{eq:QED3GNY_3d2}
\mathcal{L}^{\prime}_{3\DD} &= \sum_{i=1}^{2N_{f}}\overline{\psi}_{i}\gamma_{\mu}\left(\p_{\mu}+ieA_{\mu}\right)\psi_{i}\nonumber\\
&\quad+g\phi\sum_{i=1}^{N_f}\left(\overline{\psi}_{i}\psi_{i}-\overline{\psi}_{i+N_{f}}\psi_{i+N_{f}}\right)+\dots.
\end{align}
For comparison, in this subsection only we rewrite the Lagrangian in Eq.~(\ref{eq:QED3GNY_Lagrangian}) in terms of $N_{f}$ as follows:
\begin{equation}
\label{eq:QED3GNY_L3d}
\mathcal{L}_{3\DD}=\sum_{i=1}^{2N_{f}}\biggl[\overline{\psi}_{i}\gamma_{\mu}\left(\partial_{\mu}+ieA_{\mu}\right)\psi_{i}
+g\phi\overline{\psi}_{i}\psi_{i}\biggr]+\dots.
\end{equation}
In contrast to Eq.~(\ref{eq:QED3GNY_L3d}), the Yukawa coupling in the extrapolated $\epsilon$-expansion Lagrangian in Eq.~(\ref{eq:QED3GNY_3d2}) explicitly breaks the $\SU(2N_{f})$ flavor symmetry of pure QED$_{3}$ down to $\SU(N_{f}) \times \SU(N_{f})\times \text{U}(1)$. In the case $N_f=1$ relevant for the duality conjecture, the extrapolated Lagrangian has a Yukawa coupling of the form $\propto\phi(\overline{\psi}_1\psi_1-\overline{\psi}_2\psi_2)$, which has only a U$(1)$ flavor symmetry, whereas in the Lagrangian of Eq.~(\ref{eq:QED3GNY_Lagrangian}) the Yukawa coupling is of the form $\propto\phi(\overline{\psi}_1\psi_1+\overline{\psi}_2\psi_2)$, with full $\SU(2)$ flavor symmetry. This $\SU(2)$ flavor symmetry is important, as it forms a subgroup of the predicted emergent $\SO(5)$ symmetry at the critical point~\cite{Wang_2017}.

Thus, there are actually two variant QED$_{3}$ GNY theories: (i) the theory described by Eq.~(\ref{eq:QED3GNY_3d2}), which may be given the appellation QED$_{3}$ GNY$_{-}$ (the minus subscript denoting the sign between the two 2-component fermion bilinears)~\cite{Benvenuti2018} and (ii) the theory described by Eq.~(\ref{eq:QED3GNY_L3d}), termed QED$_{3}$ GNY$_{+}$. In terms of four-component spinors, the Yukawa interaction term for QED$_{3}$ GNY$_{+}$ is written as $i\sum_{i=1}^{N_{f}}\overline{\Psi}_{i}\widetilde{\gamma}_{3}\widetilde{\gamma}_{5}\Psi_{i}$~\cite{Kubota_2001}. 
The two QED-GNY theories can both be studied in $d=3$ dimensions, however, they are not embedded in the same $d=4$ theory. 
The QED$_{3}$ GNY$_{+}$ theory, which is the one the duality conjecture is based on and thus of interest here, is not Lorentz invariant in $d=4$ because the Yukawa coupling involves $\widetilde{\gamma}_3\widetilde{\gamma}_5$. 
To summarize, the large-$N$ calculation in strict $d=3$ accesses QED$_{3}$ GNY$_{+}$ whereas the $\epsilon$ expansion in $d=4-\epsilon$ studies QED$_{3}$ GNY$_{-}$. 
This important fact must be kept in mind when comparing results for the extrapolations of various critical exponents, and moreover this is germane to the duality conjecture which is based on QED$_{3}$ GNY$_{+}$.

One particularly difficult aspect of the $\epsilon$-expansion approach is incorporating the gamma-matrix structure in continuous $d$, while at the same time being able to reproduce the specific properties in $d=3$~\cite{Thooft_Veltman_1972,Leibbrandt_1975}. Commonly what is done in continuous $d$ calculations is to set the trace of the product of an odd number of gamma matrices to zero. This prescription was adopted in previous studies of the QED$_3$ GNY model, including not only those in the $4-\epsilon$ expansion~\cite{Janssen_He_2017,Ihrig_Janssen_2018,Zerf_Marquard_2018} but also those using the $1/N$ expansion in fixed but arbitrary $d$~\cite{GraceyQED3GNY_1992,GraceyQED3GNY_2018,AlanneBlasiQED3GNY_2018}. This procedure, however, does not account for the $d=3$ relation 
\begin{equation}
\label{eq:Gamma3}
\mathop{\mathrm{tr}}\gamma_{\mu}\gamma_{\nu}\gamma_{\lambda}=2i\epsilon_{\mu\nu\lambda},
\end{equation}
which holds for $2\times 2$ gamma matrices. (A recent attempt at incorporating this structure in the $\epsilon$-expansion method for pure-GNY models can be found in Ref.~\cite{Mihaila_Zerf_2017}.)
As will be shown in Secs.~\ref{sec:Fermion_CorrFunc}-\ref{sec:Boson_CorrFunc}, this tensor structure specific to theories with two-component Dirac fermions in $d=3$ leads to vital contributions to the scaling and anomalous dimensions, which are introduced forthwith in the next section.

\subsection{Scaling and anomalous dimensions}
\label{sec:ScalingDim}

\begin{figure*}[t]
\centering\includegraphics[width=14cm,height=2.5cm,clip]{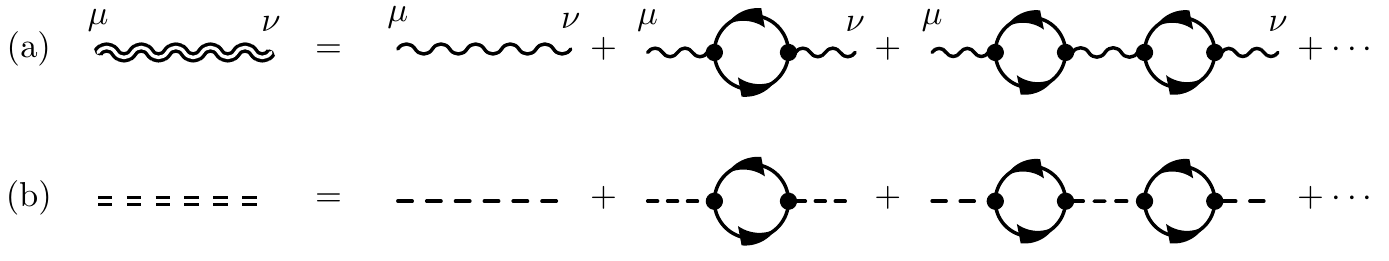}
\caption{Resummed propagators in the large-$N$ formalism for (a) the gauge field and (b) the scalar field. Solid lines denote bare fermion propagators, and single wiggly and dashed lines denote bare gauge and scalar propagators, respectively. 
Double wiggly and dashed lines represent the resummed gauge and scalar propagators, respectively.}
\label{fig:LargeN_Propagators}
\end{figure*}

In this paper we will be interested in the scaling dimensions of adjoint and singlet fermion bilinears, along with the anomalous dimension of the scalar field $\phi$ and the inverse correlation length exponent. 
To understand how these quantities are defined, consider an arbitrary local operator $\widehat{\OO}$. To first order in the $1/N$ expansion, the position-space two-point correlation function is: 
$\langle\widehat{\OO}(x)\widehat{\OO}(0)\rangle\equiv\mathcal{M}(x)=\mathcal{M}^{(0)}(x)+\frac{1}{N}\mathcal{M}^{(1)}(x)$. In the presence of an ultraviolet (UV) cutoff $\Lambda$, these functions are expected to scale as follows~\cite{Chester_Pufu_2016}:
\begin{align}
\mathcal{M}^{(0)}(x)&=\frac{A}{\left|x\right|^{2\Delta^{(0)}}},\\
\mathcal{M}^{(1)}(x)&=\frac{1}{\left|x\right|^{2\Delta^{(0)}}}\left[-B\ln(\Lambda^2\left|x\right|^2)+\mathcal{O}(\left|x\right|^0)\right].
\end{align}
The constants $A$ and $B$ are independent of $N$, and $\Delta^{(0)}$ is the scaling dimension of $\widehat{\OO}$ in the $N=\infty$ theory. 
In the large-$N$ limit, the $\mathcal{O}(1/N)$ correction can be exponentiated and thus the correlation function scales as 
\begin{equation}
\mathcal{M}(x)=\frac{\tilde{A}}{\left|x\right|^{2\Delta_{\widehat{\OO}}}},
 \end{equation}
where $\tilde{A}$ is an unimportant function of $A,B,$ and $\Lambda$. The scaling dimension for the operator $\widehat{\OO}$, to $\mathcal{O}(1/N)$, is thus
$\Delta_{\widehat{\OO}}=\Delta^{(0)}+\Delta^{(1)}\frac{1}{N}=\Delta^{(0)}+\frac{B}{A}\frac{1}{N}.$ The calculations of the correlation functions will be performed in momentum space, where $\mathcal{M}(p)\equiv\langle\widehat{\OO}(p)\widehat{\OO}(-p)\rangle$ scales as 
\begin{equation}
\mathcal{M}(p)=A\left|p\right|^{\alpha}-\frac{B}{N}\left|p\right|^{\alpha}\ln\left(\frac{\Lambda^2}{p^2}\right).
\end{equation}
Comparison with the position-space definition then gives 
\begin{equation}
\label{eq:Scaling_Dim}
\Delta_{\widehat{\OO}}=\frac{1}{2}\left(\alpha+d\right)+\frac{B}{A}\frac{1}{N}.
\end{equation}
Thus $\Delta^{(0)}=\frac{1}{2}\left(\alpha+d\right)$ and $\Delta^{(1)}=\frac{B}{A}$.

In Sec.~\ref{sec:Fermion_CorrFunc}, two gauge-invariant operators will be considered: the $\SU(N)$ flavor-adjoint mass operator $\widehat{\OO}=\frac{1}{\sqrt{N}}\overline{\psi}T_{A}\psi$ 
and the $\SU(N)$ flavor-singlet mass operator $\widehat{\OO}=\frac{1}{\sqrt{N}}\overline{\psi}\psi$.  Here, $T_{A}$ denotes the generators of $\SU(N)$, which are traceless, Hermitian $N\times N$ matrices. In the duality conjecture~\cite{Wang_2017}, the bilinear whose scaling dimension should relate to $\nu_\text{N\'eel-VBS}^{-1}$ via Eq.~(\ref{eq:And1}) is $\overline{\psi}_1\psi_1-\overline{\psi}_2\psi_2$, which transforms in the adjoint of $\SU(2)$.
Both adjoint and singlet bilinear scaling dimensions will be obtained from an analysis of the fermion four-point correlation function. Similarly, in Sec.~\ref{sec:Boson_CorrFunc} the scaling dimensions of the bosonic operators $\widehat{\OO}=\phi,\phi^2$ will be computed, 
via studying the bosonic two- and four-point correlation functions, respectively. From the scaling dimension $\Delta_{\phi^{2}}$, the inverse correlation length exponent, $\nu^{-1}$, is then defined as $\nu^{-1}=d-\Delta_{\phi^{2}}$. 
The other quantity of interest is the bosonic anomalous dimension, $\eta_{\phi}$, which arises by considering the scaling of the bosonic two-point function: 
\begin{equation}
\label{eq:eta}
G_{\phi}\left(p\right)\equiv\langle\phi(p)\phi(-p)\rangle\sim\frac{1}{\left|p\right|^{2-\eta_{\phi}}}.
\end{equation}
Unitarity bounds in conformal field theory~\cite{Ferrara_1974,Mack_1977} impose constraints on these scaling and anomalous dimensions. 
In particular, the scaling dimension of a Lorentz scalar $\Delta$ must obey $\Delta\geq \frac{d}{2}-1$. Since $\Delta_{\phi}=(d-2+\eta_{\phi})/2$ and $\Delta_{\phi^{2}}=d-\nu^{-1}$, this implies that $\eta_{\phi}\geq0$ and $\nu^{-1}\leq1+d/2$.

\begin{table}
\caption{$1/N$-expansion results for QED$_{3}$ GNY critical exponents. Approximants that either do not exist or are negative are labeled $\times$.}
\begin{tabular}{|c|c|c|c|c|}
\hline 
 & Analytical result & $N=2$ & $\Pade$ & Borel-$\Pade$\tabularnewline
\hline 
\hline 
$\nu^{-1}$ & $1+80/(\pi^{2}N)$ & $5.053$ & $\times$ & $\times$\tabularnewline
\hline 
$\eta_{\phi}$ & $1-96/(\pi^{2}N)$ & $-3.863$ & $0.1705$ & $0.3031$\tabularnewline
\hline 
$\Delta_{\phi}$ & $1-48/(\pi^{2}N)$ & $-1.432$ & $0.2914$ & $0.4243$\tabularnewline
\hline 
$\Delta_{\overline{\psi}T_{A}\psi}$ & $2-16/(\pi^{2}N)$ & $1.189$ & $1.423$ & $1.514$\tabularnewline
\hline 
\end{tabular}\label{tab:Results}
\end{table}

A summary of our $1/N$-expansion results is presented in Table~\ref{tab:Results}. In order to extrapolate these results to the $N=2$ case, 
applicable for the equalities conjectured in Eqs.~(\ref{eq:Eta1}-\ref{eq:And1}), we apply $\Pade$ and Borel-$\Pade$ approximants (see Appendix~\ref{sec:PBP}). 
Numerical studies~\cite{Sandvik_2007,Melko_2008,Nahum_2015} of the anomalous dimensions $\eta_{\text{N\'eel}}$ and $\eta_{\VBS}$ for the N\'eel and VBS order parameters give values in the range $\sim0.25-0.35$,  
while the $\Pade$ and Borel-$\Pade$ approximants obtained in this paper give the result
\begin{equation}
\eta_{\phi}\approx0.17-0.30.
\end{equation}
Numerical results~\cite{Sandvik_2007,Melko_2008,Nahum_2015} for the inverse correlation length exponent $\nu_\text{N\'eel-VBS}^{-1}$ are in the range $\sim1.3-2$. From Table~\ref{tab:Results}, the $\Pade$ and Borel-$\Pade$ approximants for the adjoint bilinear scaling dimension yield the following estimate:
\begin{equation}
3-\Delta_{\overline{\psi}T_{A}\psi}\approx1.49-1.58.
\end{equation}
The reasonable concurrence between these estimates and numerical results for the N\'eel-to-VBS transition supports the predicted equalities Eq.~(\ref{eq:Eta1}-\ref{eq:And1}) and gives credence to the duality conjecture. 
Unfortunately, while the QED$_3$ GNY exponent $\nu^{-1}$ is positive for all $N$ at $\mathcal{O}(1/N)$, when $N=2$ the $\mathcal{O}(1/N)$ result gives a value outside the unitarity bound in $d=3$. Moreover, neither of the approximants are applicable. 
Thus, at this order a direct comparison between the value of $\nu^{-1}$ in the QED$_3$ GNY theory and $3-\Delta_{\overline{\psi}T_{A}\psi}$ is unfeasible. 
In the remaining sections we present the derivations of the results appearing in Table~\ref{tab:Results}. 

\begin{figure}[b]
\centering\includegraphics[width=8cm,height=1cm,clip]{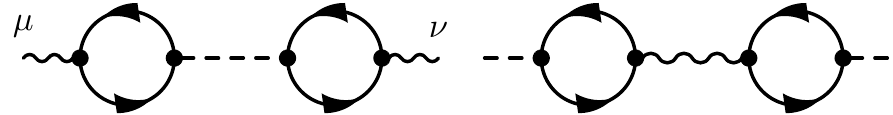}
\caption{Vanishing contributions to the gauge and scalar propagators.}
\label{fig:Propagator_Vanish}
\end{figure} 

\subsection{Large-$N$ Feynman rules}

\begin{figure*}[t]
\centering\includegraphics[width=15cm,height=3cm,clip]{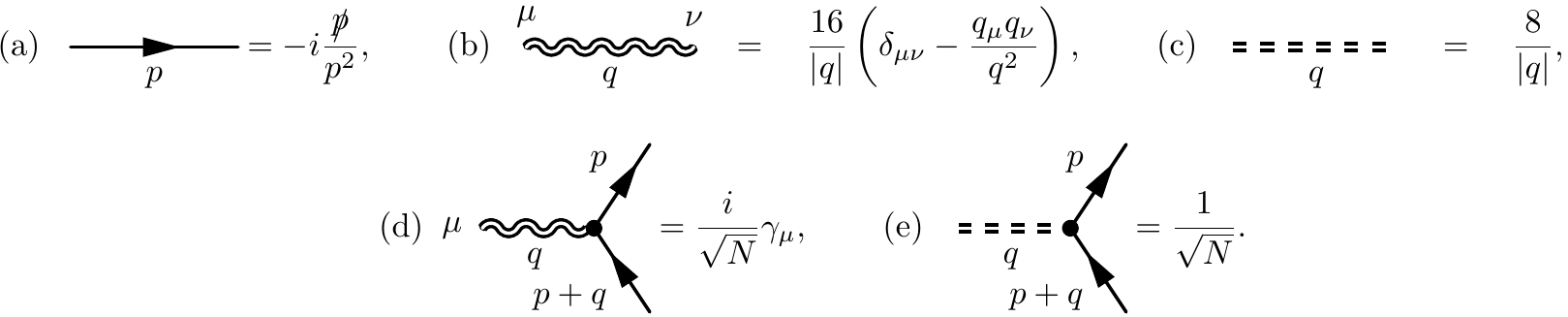}
\caption{Propagators and vertices in the large-$N$ formalism. (a) The bare fermion propagator; (b) and (c) the large-$N$ gauge and scalar propagators, respectively; 
(d) and (e) the large-$N$ vertices for the gauge and scalar interactions with the bare fermions, respectively.}
\label{fig:LargeN_FeynmanRules}
\end{figure*}

Using the large-$N$ formalism it can be shown that, as in the pure QED$_{3}$ and GNY models, the QED$_3$ GNY model flows to an interacting conformal field theory in the infrared (IR) limit. Mathematically this limit corresponds to momentum scales $q$ that are small in comparison to the dimensionful couplings $e^{2}$ and $g^{2}$: $|q|\ll e^{2},g^{2}$. 
From Eqs.~(\ref{eq:Scalar_prop}-\ref{eq:Gauge_prop}), in this limit the scalar and gauge propagators are dominated by self-energy diagrams with fermion-loop corrections.
For the fermion propagator, however, the bare propagator in Eq.~(\ref{eq:Fermion_prop}) will suffice. 
As discussed in Ref.~\cite{Chester_Pufu_2016}, it is not possible to first take the limit $e^2\gg|q|$, and then resum the gauge propagator corrections, except in the Landau gauge. 
To circumvent this problem Ref.~\cite{Chester_Pufu_2016} used a one-parameter family of nonlocal gauge-fixing terms~\cite{Gracey_2000} allowing this limit to be performed from the start.

The one-particle irreducible (1PI) self-energy of the gauge field consists of a closed fermion loop, and by summing the full geometric series of such terms the resummed propagator is obtained. 
The diagrammatic expansion is indicated in Fig.~\ref{fig:LargeN_Propagators}(a). The result for the gauge self-energy, in $d=3$ dimensions, is
\begin{equation}
\Sigma^{(0)}_{\mu\nu}\left(q\right)=-\left(\delta_{\mu\nu}-\frac{q_{\mu}q_{\nu}}{q^{2}}\right)\frac{\left|q\right|}{16}.
\end{equation}
From Dyson's equation the resummed propagator is then
\begin{align}
\Pi_{\mu\nu}\left(q\right)&=\left\{ \left[\Pi^{0}\left(q\right)\right]^{-1}-\Sigma^{(0)}\left(q\right)\right\}^{-1}_{\mu\nu},\no\\
&=\frac{16}{\left|q\right|}\left(\delta_{\mu\nu}-\frac{q_{\mu}q_{\nu}}{q^{2}}\right)\left[1+\mathcal{O}\left(\frac{\left|q\right|}{e^{2}}\right)\right].
\end{align}
In the last step we have set $\xi=0$, i.e., taken the Landau-gauge limit. As previously mentioned, the higher-order terms are suppressed in the IR limit. 
The resummed gauge propagator thus obeys $q_{\mu}\Pi_{\mu\nu}\left(q\right)=0$; this result corresponds to a nonlocal gauge-fixing parameter $\zeta=1$ in the notation of Ref.~\cite{Chester_Pufu_2016}.

The 1PI scalar self-energy again consists of a closed fermion loop, and the diagrammatic expansion is indicated in Fig.~\ref{fig:LargeN_Propagators}(b). 
In $d=3$ dimensions, the scalar self-energy is 
\begin{equation}
\Sigma^{(0)}_{\phi}\left(q\right)=-\frac{1}{8}\left|q\right|.
\end{equation}
Using Dyson's equation, the resummed propagator is
\begin{align}
D_{\phi}\left(q\right)&=\left\{ \left[G_{\phi}^{0}\left(q\right)\right]^{-1}-\Sigma^{(0)}_{\phi}\left(q\right)\right\} ^{-1},\no\\
&=\frac{8}{\left|q\right|}\left[1+\mathcal{O}\left(\frac{\left|q\right|}{g^{2}}\right)\right].
\end{align}
Finally, note that the above calculations of the gauge and scalar propagators do not incorporate any $\mathcal{O}(N^{0})$ mixing between the two, namely, no Yukawa vertices appear in the gauge self-energy $\Sigma^{(0)}_{\mu\nu}$ and no QED vertices appear in the scalar self-energy $\Sigma^{(0)}_{\phi}$. Indeed, all such terms are zero. In Fig.~\ref{fig:Propagator_Vanish} a class of these vanishing diagrams is shown. 
 
The vertices connecting the resummed propagators with the bare fermions are the same as those given in Eqs.~(\ref{eq:Scalar_vertex}) and (\ref{eq:Gauge_vertex}). 
Thus, the complete Feynman rules in the large-$N$ formalism, for the propagators and the vertices, are completely specified and given in Fig.~\ref{fig:LargeN_FeynmanRules}. 
In the proceeding sections these Feynman rules will be used to determine the scaling and anomalous dimensions.

\section{Fermion four-point correlation function}
\label{sec:Fermion_CorrFunc}

\begin{figure}[h]
\centering\includegraphics[width=3cm,height=2cm,clip]{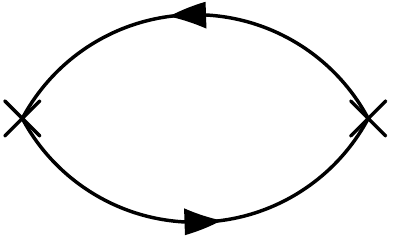}
\caption{Lowest-order [$\mathcal{O}(N^{0})$] contribution to the fermion four-point correlation function. The $\times$ denotes the prefactor $1/\sqrt{N}$.}
\label{fig:Fermion_Bubble}
\end{figure}

In this section the scaling dimensions of the adjoint and singlet fermion mass operators are computed to $\mathcal{O}(1/N)$. 
The calculation is separated into three subsections delineated as follows: Sec.~\ref{sec:QED3} considers pure QED$_{3}$, while in Sec.~\ref{sec:GNY} pure GNY is studied, and 
finally Sec.~\ref{sec:QED3_GNY} analyzes QED$_{3}$ GNY and calculates the pertinent $\mathcal{O}(1/N)$ diagrams containing both gauge and scalar fields. 
The  $\mathcal{O}(N^{0})$ contribution to the fermion four-point correlation function is shown in Fig.~\ref{fig:Fermion_Bubble}. For external momentum $p$, this diagram is 
\begin{equation}
\label{eq:Fermion_Bubble}
\mathcal{M}(p)_{4}=-\frac{|p|}{8}.
\end{equation}

\subsection{QED$_{3}$}
\label{sec:QED3}

\begin{figure}[t]
\centering\includegraphics[width=8.75cm,height=4.75cm,clip]{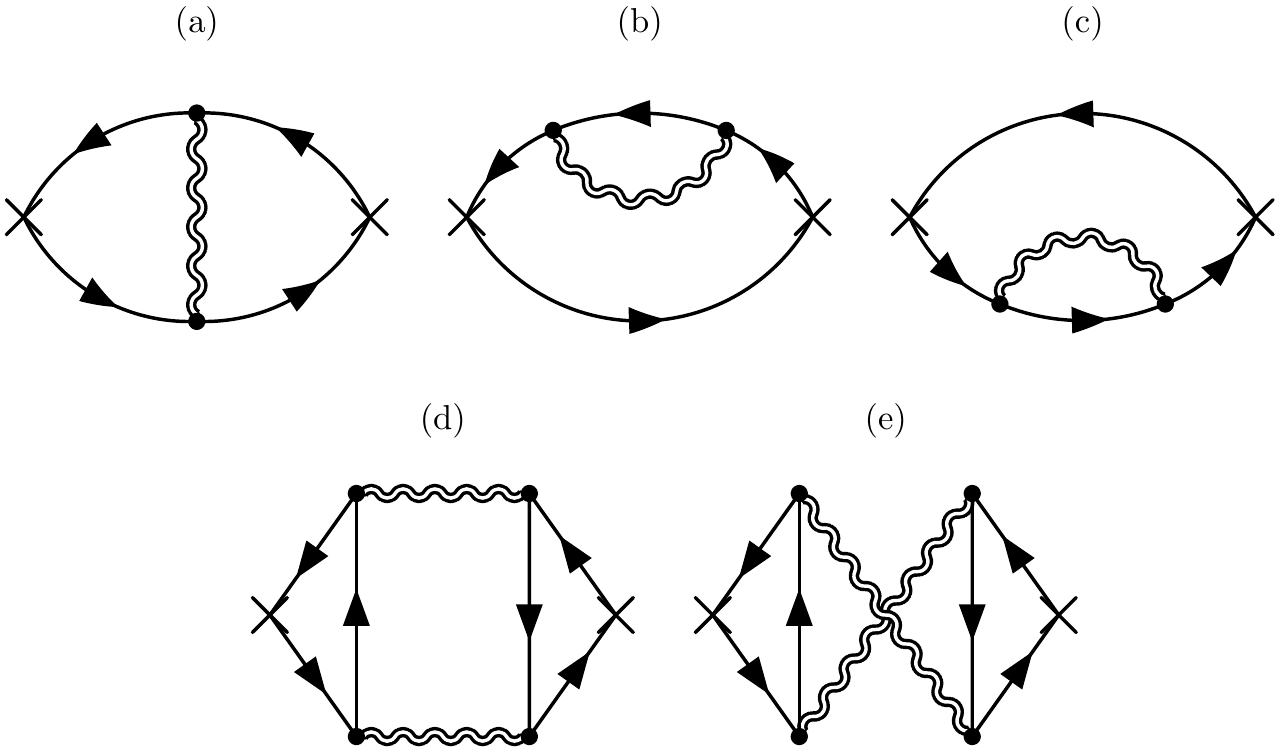}
\caption{The $\mathcal{O}(1/N)$ corrections to the fermion four-point correlation function in pure QED$_{3}$. The Aslamazov-Larkin diagrams in (d) and (e) vanish for the adjoint bilinear.}
\label{fig:QED3_Diagrams}
\end{figure} 

Large-$N$ studies of QED$_{3}$ abound in the literature. In particular, the $\mathcal{O}(1/N)$ corrections to the fermion four-point correlation function are well known and consist of the five diagrams shown in Fig.~\ref{fig:QED3_Diagrams}. 
The three diagrams in Figs.~\ref{fig:QED3_Diagrams}(a-c) were computed in Ref.~\cite{Rantner_Wen_2002}, and we have independently verified their results. 
If $p$ denotes the external momentum and $\Lambda$ the UV cutoff, then their logarithmically singular part is 
\begin{equation}
\label{eq:QED3_ABC}
\mathcal{M}(p)_{5\A+5\B+5\C}=-\frac{8|p|}{3\pi^2N}\ln\left(\frac{\Lambda^2}{p^2}\right).
\end{equation}
 
The two QED$_{3}$ AL diagrams appearing in Figs.~\ref{fig:QED3_Diagrams}(d-e) were first (indirectly) considered in Ref.~\cite{Hermele_Erratum_2007}, via fermion self-energy corrections, and subsequently they were computed explicitly in Ref.~\cite{Chester_Pufu_2016}, the latter which we have again independently verified. Furthermore, Refs.~\cite{Hermele_Erratum_2007,Chester_Pufu_2016} proved that they contribute to only the singlet bilinear scaling dimension; 
this is because the trace of the adjoint generator, $T_{A}$, which arises in the computation of the triangle vertex in Figs.~\ref{fig:QED3_Diagrams}(d-e), is zero. The two AL diagrams give equal contributions and their combined logarithmically singular part is
\begin{equation}
\label{eq:QED3_DE}
\mathcal{M}(p)_{5\D+5\E}=\frac{8|p|}{\pi^2N}\ln\left(\frac{\Lambda^2}{p^2}\right).
\end{equation}

Combining Eqs.~(\ref{eq:Fermion_Bubble}-\ref{eq:QED3_DE}) and using Eq.~(\ref{eq:Scaling_Dim}), the adjoint and singlet fermion bilinear scaling dimensions in QED$_{3}$ are then
\begin{align}
\Delta^{\text{QED}_{3}}_{\overline{\psi}T_{A}\psi}=2-\frac{64}{3\pi^{2}N},\\
\Delta^{\text{QED}_{3}}_{\overline{\psi}\psi}=2+\frac{128}{3\pi^{2}N}.
\end{align}
These two results agree with Eq.~(2.20) and Eq.~(2.19) in Ref.~\cite{Chester_Pufu_2016}, respectively. A large-$N$ analysis in arbitrary dimension $d$ was performed in Ref.~\cite{GraceyQED3_1993},
and their scaling dimension to $\mathcal{O}(1/N)$, which can be inferred from their Eq.~(27), agrees with the adjoint scaling dimension given here.
However, it is important to note that Ref.~\cite{GraceyQED3_1993} worked in arbitrary $d$ and assumed that the trace of an odd number of gamma matrices vanishes, which disregards Eq.~(\ref{eq:Gamma3}). 
Thus, the contribution from the QED$_{3}$ AL diagrams did not arise in that reference. As these diagrams contribute to only the singlet scaling dimension, this explains why in fixed $d = 3$ 
the adjoint and singlet scaling dimensions do not coincide at $\mathcal{O}(1/N)$.

\subsection{GNY}
\label{sec:GNY}

\begin{figure}[t]
\centering\includegraphics[width=8.75cm,height=4.75cm,clip]{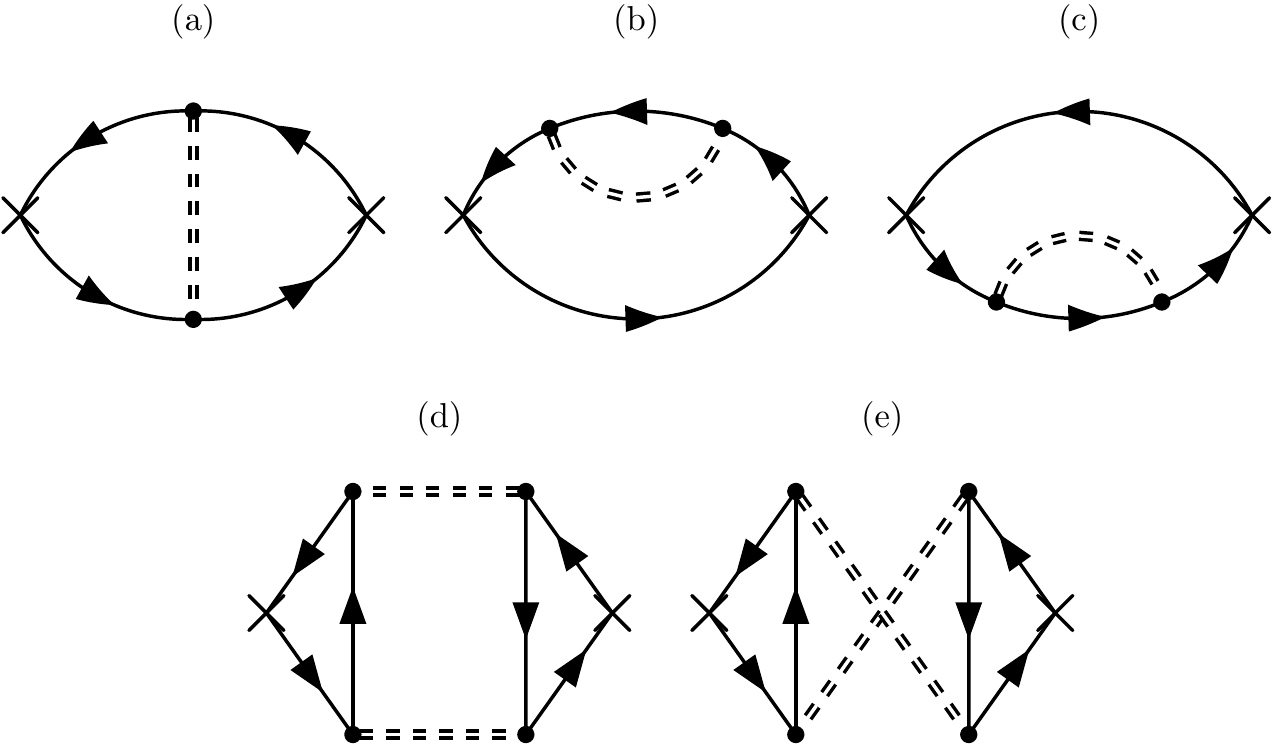}
\caption{The $\mathcal{O}(1/N)$ corrections to the fermion four-point correlation function in pure GNY. The Aslamazov-Larkin diagrams in (d) and (e) vanish identically.}
\label{fig:GNY_Diagrams}
\end{figure} 

The fermion four-point correlation function corrections in GNY are given by the same diagrams as in QED$_{3}$, but with a scalar field replacing the gauge field. 
Only the final results are presented here, while a more detailed derivation is deferred to Appendix~\ref{app:GNY}. 
The logarithmically singular part of the three diagrams in Figs.~\ref{fig:GNY_Diagrams}(a-c) is 
\begin{equation}
\label{eq:GNY_ABC}
\mathcal{M}(p)_{6\A+6\B+6\C}=\frac{2|p|}{3\pi^2N}\ln\left(\frac{\Lambda^2}{p^2}\right).
\end{equation}
This result agrees with the calculation of Ref.~\cite{Hands_1991}, which can be deduced from their Eq.~(4.5). 
The two GNY AL diagrams in Figs.~\ref{fig:GNY_Diagrams}(d-e) give equal contributions. Nonetheless, the triangle vertex appearing in these diagrams is zero and so they vanish:
\begin{equation}
\label{eq:GNY_DE}
\mathcal{M}(p)_{6\D+6\E}=0.
\end{equation}
Therefore, the adjoint and singlet scaling dimensions in GNY are equal at $\mathcal{O}(1/N)$. 

Combining Eqs.~(\ref{eq:Fermion_Bubble},\ref{eq:GNY_ABC},\ref{eq:GNY_DE}) and using Eq.~(\ref{eq:Scaling_Dim}), the adjoint and singlet scaling dimensions in GNY are then~\cite{Comment1}
\begin{align}
\Delta^{\text{GNY}}_{\overline{\psi}T_{A}\psi}=2+\frac{16}{3\pi^{2}N},\\
\Delta^{\text{GNY}}_{\overline{\psi}\psi}=2+\frac{16}{3\pi^{2}N}.
\end{align}
The adjoint scaling dimension has been computed in fixed $d = 3$ dimensions for GNY in Eq.~(B8) of Ref.~\cite{Iliesiu_2016}, and their result agrees with that given here. 
The scaling dimension can also be inferred from the results of Ref.~\cite{GraceyGN_1992}, which considered a large-$N$ expansion for arbitrary dimension $d$ but without accounting for Eq.~(\ref{eq:Gamma3}). 
The result found in that reference agrees with the one given here, however, it should be noted that this is because the scalar-scalar AL diagrams give no contribution to the fermion four-point correlation function.

\subsection{QED$_{3}$ GNY}
\label{sec:QED3_GNY}

\begin{figure}[t]
\centering\includegraphics[width=7cm,height=6cm,clip]{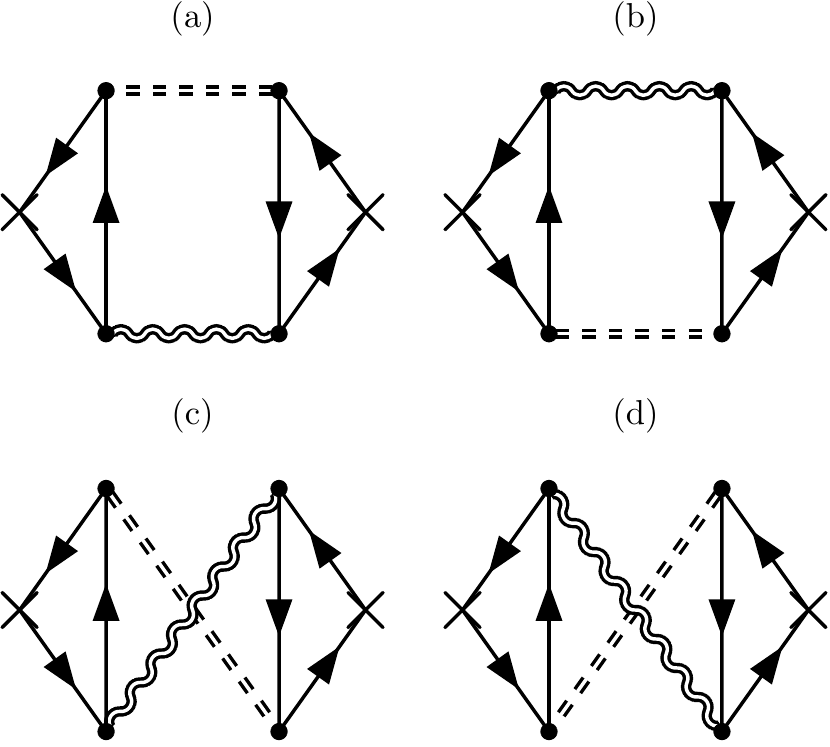}
\caption{The $\mathcal{O}(1/N)$ corrections to the fermion four-point correlation function with gauge and scalar propagators.}
\label{fig:QED3GNY_Diagrams}
\end{figure} 

The only diagrams in QED$_{3}$ GNY that potentially contribute to the fermion scaling dimensions at $\mathcal{O}(1/N)$ are the four gauge-scalar AL diagrams shown in Figs.~\ref{fig:QED3GNY_Diagrams}(a-d). 
As shown in Appendix~\ref{app:QED3_GNY}, the exact result for Fig.~\ref{fig:QED3GNY_Diagrams}(a) is
\begin{equation}
\label{eq:QED3GNY_A}
\mathcal{M}(p)_{7\A}=\frac{2\left|p\right|}{3N}\left(1-\frac{9}{\pi^{2}}\right).
\end{equation}
It is important to emphasize that the above result is specific to the gauge where $q_{\mu}\Pi_{\mu\nu}(q)=0$, that is, the $\zeta=1$ gauge in the notation of Ref.~\cite{Chester_Pufu_2016}. 
In this particular gauge the potentially logarithmically singular term in the gauge-scalar AL diagrams is purely longitudinal in the internal photon momentum, and thus by the Ward identity it vanishes. 
Similar phenomena have arisen in other contexts~\cite{Huh_Strack_2013,Huh_Strack_2015}. 
In an arbitrary gauge, however, the gauge-scalar AL diagrams do have a logarithmic singularity, proportional to $(\zeta-1)$; we omit the details. 
The QED$_{3}$ diagrams themselves are gauge invariant~\cite{Chester_Pufu_2016}, and so by gauge invariance this gauge dependence must cancel out. 
Indeed, the diagrams in Fig.~\ref{fig:QED3GNY_Diagrams} are related to one another by $\text{7(a)} = \text{7(b)}$, $\text{7(c)} = \text{7(d)}$, $\text{7(a)} = -\text{7(c)}$, and $\text{7(b)} = -\text{7(d)}$. 
Thus, the total contribution from the QED$_{3}$ GNY AL diagrams is zero and gauge invariance is satisfied:
\begin{equation}
\label{eq:QED3GNY}
\mathcal{M}(p)_{7\A+7\B+7\C+7\D}=0.
\end{equation}
Similar types of cancellations are exhibited in Ref.~\cite{Thomson_Sachdev_2017}.

The total $\mathcal{O}(1/N)$ correction to the fermion four-point correlation function is therefore found from adding together those of pure QED$_{3}$ [Eqs.~(\ref{eq:QED3_ABC},\ref{eq:QED3_DE})] 
and pure GNY [Eqs.~(\ref{eq:GNY_ABC},\ref{eq:GNY_DE})]. 
Thus, the QED$_{3}$ GNY fermion bilinear scaling dimensions are~\cite{Comment1}
\begin{align}
\Delta^{\text{QED}_{3} \text{GNY}}_{\overline{\psi}T_{A}\psi}=2-\frac{16}{\pi^{2}N},\\
\label{eq:QED3GNY_FermSingDim}
\Delta^{\text{QED}_{3} \text{GNY}}_{\overline{\psi}\psi}=2+\frac{48}{\pi^{2}N}.
\end{align} 
The singlet scaling dimension can be compared to the corresponding result in the $\mathcal{O}(1/N)$ calculations of Ref.~\cite{GraceyQED3GNY_1992} with four-component Dirac fermions, which in fact agrees with the adjoint scaling dimension calculated here. 
More recently, the adjoint and singlet scaling dimensions for four-component fermions are given at $\mathcal{O}(1/N^2)$ in Eqs.~(4.4) and (4.6), respectively, of Ref.~\cite{GraceyQED3GNY_2018}. 
The results given there, to $\mathcal{O}(1/N)$, agree with one another and are equal to the adjoint scaling dimension calculated here. 
The reason for these discrepancies was discussed already in regard to pure QED$_{3}$, where a cogent argument was that the QED$_{3}$ AL diagrams are not included in the previous arbitrary-$d$, large-$N$ studies, and moreover they contribute to only the singlet scaling dimension with two-component fermions. As discussed in Sec.~\ref{sec:diracalgebra}, the four-component SU$(N_f)$ singlet bilinear $\sum_{i=1}^{N_f}\overline{\Psi}_i\Psi_i$, whose scaling dimension is calculated in Ref.~\cite{GraceyQED3GNY_1992,GraceyQED3GNY_2018}, 
is equal to $\sum_{i=1}^{N_f}\left(\overline{\psi}_i\psi_i-\overline{\psi}_{i+N_f}\psi_{i+N_f}\right)$ in terms of two-component fermions, which transforms in the adjoint of SU$(N)$ where $N=2N_f$.

\section{Scalar correlation functions}
\label{sec:Boson_CorrFunc}

\subsection{Two-point correlation function}

\begin{figure}[t]
\centering\includegraphics[width=8.5cm,height=1.5cm,clip]{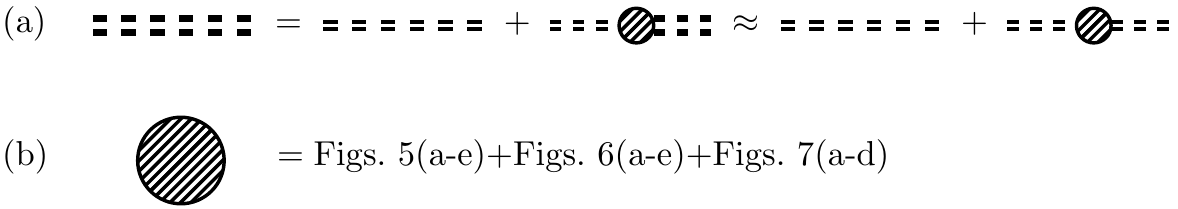}
\caption{(a) The Dyson equation and its $\mathcal{O}\left(1/N\right)$ solution for the bosonic two-point correlation function $G_{\phi}$. (b) Scalar self-energy diagrams that contribute at $\mathcal{O}(1/N)$.}
\label{fig:Boson_TwoPoint_Function}
\end{figure}

In this section we compute the scaling dimensions of the scalar $\phi$ and $\phi^2$ operators from the two and four-point correlation functions of $\phi$, respectively. The calculation of these quantities can be performed simultaneously for GNY and QED$_{3}$ GNY, 
thus it is expedient to employ a general notation that encompasses both theories. The scalar two-point function is denoted by $G_{\phi}\left(p\right)\equiv\langle\phi(p)\phi(-p)\rangle$, and its
$\mathcal{O}\left(1/N\right)$ 1PI scalar self-energy contribution is represented by $\Sigma^{(1)}_{\phi}\left(p\right)$. From Dyson's equation the two-point function to $\mathcal{O}(1/N)$ is 
\begin{align}
\label{eq:Gamma2}
G_{\phi}\left(p\right) &= D_{\phi}\left(p\right)+D_{\phi}\left(p\right)\Sigma^{(1)}_{\phi}\left(p\right)G_{\phi}\left(p\right) ,\nonumber\\
& \approx D_{\phi}\left(p\right)+D_{\phi}\left(p\right)\Sigma^{(1)}_{\phi}\left(p\right)D_{\phi}\left(p\right).
\end{align}
The Feynman diagram for this equation is shown in Fig.~\ref{fig:Boson_TwoPoint_Function}(a). 
The shaded bubble in Fig.~\ref{fig:Boson_TwoPoint_Function}(a) corresponds to the insertion of $\Sigma^{(1)}_{\phi}(p)$; that is, it is the sum of diagrams in Figs.~\ref{fig:QED3_Diagrams}(a-e), Figs.~\ref{fig:GNY_Diagrams}(a-e), and Figs.~\ref{fig:QED3GNY_Diagrams}(a-d),
as indicated in Fig.~\ref{fig:Boson_TwoPoint_Function}(b). 

The self-energy $\Sigma^{(1)}_{\phi}$ has the following generic form:
\begin{equation}
\label{eq:1PI}
\Sigma^{(1)}_{\phi}\left(p\right)=\frac{c\left|p\right|}{\pi^{2}N}\ln\left(\frac{\Lambda^{2}}{p^{2}}\right).
\end{equation}
The coefficient $c$ here is not arbitrary, it is a theory-specific constant; for instance, from Eqs.~(\ref{eq:GNY_ABC},\ref{eq:GNY_DE}) it follows that $c_{\text{GNY}}=\frac{2}{3}$ for pure GNY, whereas from Eqs.~(\ref{eq:QED3_ABC},\ref{eq:QED3_DE},\ref{eq:GNY_ABC},\ref{eq:GNY_DE}),  
$c_{\text{QED}_{3}\text{-GNY}}=6$ for QED$_{3}$ GNY. On the other hand, if the QED$_{3}$ AL diagrams in Eq.~(\ref{eq:QED3_DE}) are excluded, then the constant is $c^{\prime}_{\text{QED}_{3}\text{-GNY}}=-2$ (no AL). 

Using Eq.~(\ref{eq:eta}) and Eqs.~(\ref{eq:Gamma2}-\ref{eq:1PI}), the anomalous dimension of $\phi$ is then
\begin{equation}
\eta_{\phi}=1-\frac{16c}{\pi^{2}N}.
\end{equation}
For the cases of GNY and QED$_{3}$ GNY, this becomes
\begin{align}
\label{eq:eta_phi}
\eta_{\phi} = \left\{\begin{array}{lll}
1-32/(3\pi^{2}N), & \text{GNY}, \\
1+32/(\pi^{2}N), & \text{QED}_{3}\text{-GNY}\ \text{(no AL)},  \\ 
1-96/(\pi^{2}N), & \text{QED}_{3}\text{-GNY}.\end{array}\right.
\end{align}
The anomalous dimension for GNY agrees with the well-known result~\cite{ZinnJustin_1991,Hands_1991,GraceyGN_1992,ZinnJustin_2003}.
Similarly, the result for QED$_{3}$ GNY, with the exclusion of the QED$_{3}$ AL diagrams, agrees with that of Ref.~\cite{GraceyQED3GNY_1992}. 
However, in fixed $d=3$ dimensions the AL diagrams give a nonzero contribution, as shown in the third line of Eq.~(\ref{eq:eta_phi}).  

The scalar-field scaling dimension $\Delta_{\phi}$ can be determined from the anomalous dimension via $\Delta_{\phi}=\left(d-2+\eta_{\phi}\right)/2$. 
From the results in Eq.~(\ref{eq:eta_phi}), we then have
\begin{align}
\Delta_{\phi} = \left\{\begin{array}{lll}
 1-16/(3\pi^{2}N), & \text{GNY}, \\
 1+16/(\pi^{2}N), & \text{QED}_{3}\text{-GNY\ (no AL)},  \\ 
 1-48/(\pi^{2}N), & \text{QED}_{3}\text{-GNY}.\end{array}\right.
\end{align}
The pure-GNY result agrees with Refs.~\cite{Hands_1991,Rosenstein_1991,Iliesiu_2016,Iliesiu_2018,ZinnJustin_2003}. Adding the result in the last line above to that of Eq.~(\ref{eq:QED3GNY_FermSingDim}) then gives
$\Delta_{\overline{\psi}\psi}+\Delta_{\phi}=3$, in QED$_{3}$ GNY. 

\subsection{Four-point correlation function}

\begin{figure}[h]
\centering\includegraphics[width=8.5cm,height=2.25cm,clip]{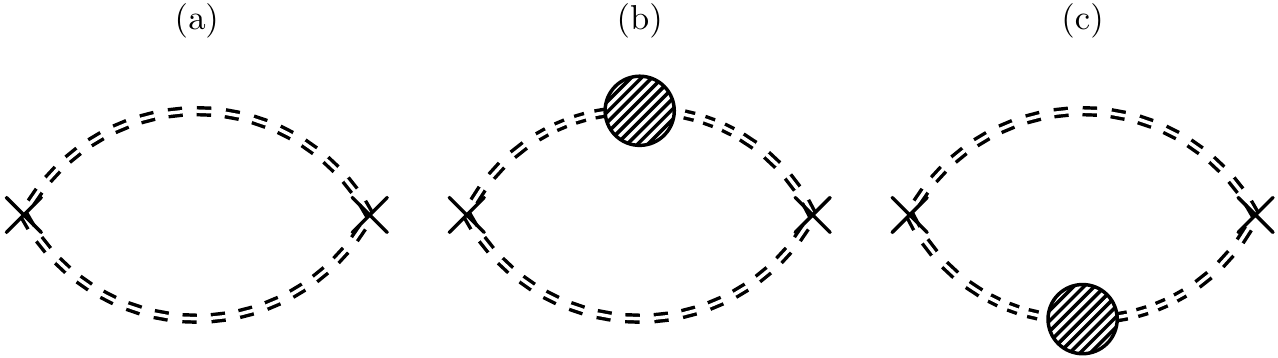}
\caption{Scalar bubble contributions to the scalar four-point function.}
\label{fig:GNY_Boson_Diagram}
\end{figure} 

\begin{figure}[t]
\centering\includegraphics[width=6cm,height=7cm,clip]{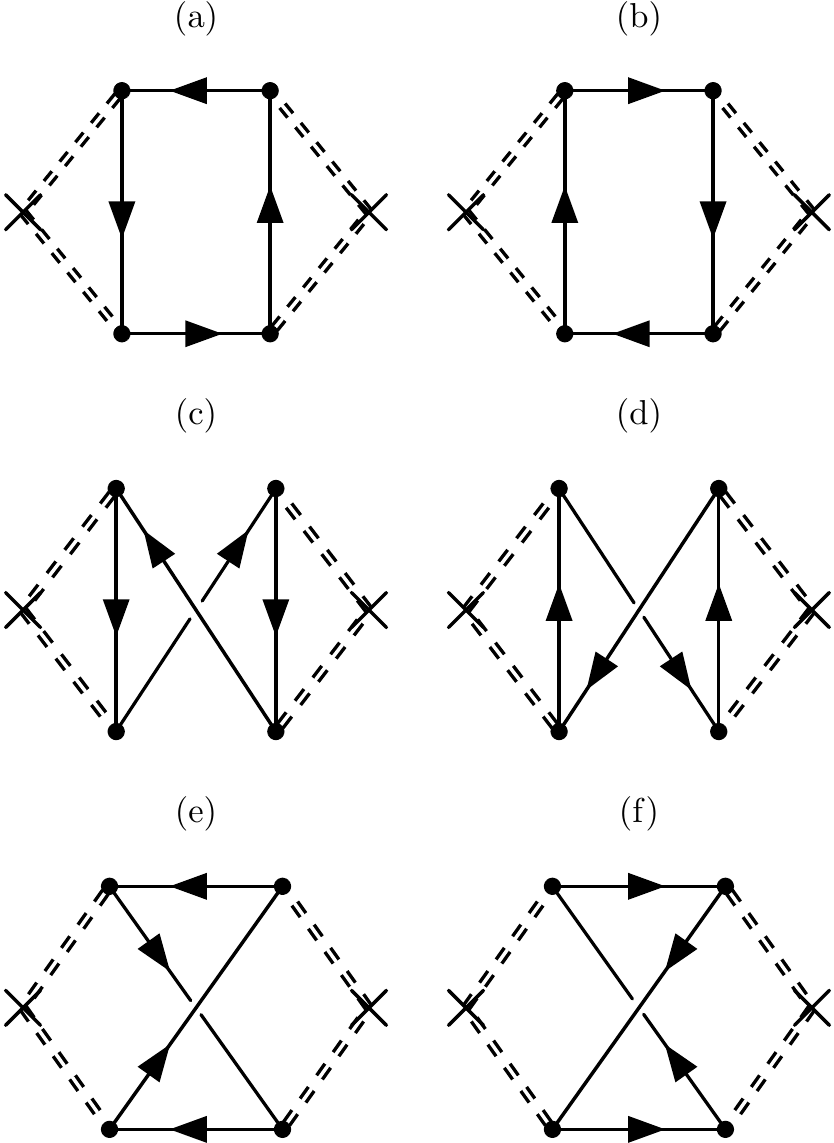}
\caption{Fermion box contributions to the scalar four-point function.}
\label{fig:Box_BowTie_HourGlass_Diagrams}
\end{figure} 

To determine the scaling dimension $\Delta_{\phi^2}$ of the $\phi^2$ operator, we now investigate the scalar four-point correlation function, to be denoted by $G_{\phi^{2}}\left(p\right)\equiv\langle\phi^2(p)\phi^2(-p)\rangle$. 
The lowest order contribution is shown in Fig.~\ref{fig:GNY_Boson_Diagram}(a), and evaluating this diagram gives
\begin{equation}
\label{eq:Boson_Bubble}
G_{\phi^{2},9\A}(p) = 2\int\frac{d^{3}k}{\left(2\pi\right)^{3}}\frac{8^2}{\left|k\right|\left|k+p\right|} = -\frac{2\times8^{2}}{\left(2\pi\right)^{2}}\left|p\right|.
\end{equation}
In this calculation a linearly divergent term has been ignored, since it does not contribute to the scaling dimension. Note that, the prefactor of 2 appears here because there are two such diagrams, as can be verified using Wick's theorem. 
The $\mathcal{O}(1/N)$ corrections to the scalar-scalar bubble diagram are shown in Figs.~\ref{fig:GNY_Boson_Diagram}(b,c). Using the constant $c$ facilitates calculating these diagrams for both the pure GNY and QED$_{3}$ GNY theories all at once.
The result is
\begin{align}
G_{\phi^{2},9\B+9\C}(p) &=4\times8^{3}\int\frac{d^{3}k}{\left(2\pi\right)^{3}}\frac{\Sigma^{(1)}_{\phi}\left(k\right)}{k^{2}\left|k-p\right|},\nonumber\\
&=-\frac{32\times8^2c}{\pi^{2}N}\frac{\left|p\right|}{\left(2\pi\right)^{2}}\ln\left(\frac{\Lambda^{2}}{p^{2}}\right),\nonumber\\
&=G_{\phi^{2},9\A}(p)\frac{16c}{\pi^{2}N}\ln\left(\frac{\Lambda^{2}}{p^{2}}\right).
\end{align}
A linearly divergent term has again been dropped, and only the logarithmically singular part has been retained. 

The remaining $\mathcal{O}(1/N)$ corrections to the scalar four-point correlation function consist of diagrams with an internal fermion ``box", as shown in Figs.~\ref{fig:Box_BowTie_HourGlass_Diagrams}(a-f). 
The calculation of these diagrams is deferred to Appendix~\ref{app:BoxFunctions}, and the final result is that the fermion box function contribution is 
\begin{equation}
\label{eq:Fermion_Box}
G_{\phi^{2},\text{box}}(p)=-G_{\phi^{2},9\A}(p)\frac{16}{\pi^{2}N}\ln\left(\frac{\Lambda^{2}}{p^{2}}\right).
\end{equation}
The total scalar four-point correlation function, ignoring unimportant linearly divergent terms, is thus
\begin{equation}
G_{\phi^{2}}(p)=G_{\phi^{2},9\A}(p)\left[1+\frac{16\left(c-1\right)}{\pi^{2}N}\ln\left(\frac{\Lambda^{2}}{p^{2}}\right)\right].
\end{equation}
Thus, the scaling dimension of $\phi^2$ is 
\begin{equation}
\Delta_{\phi^{2}}=2+\frac{16\left(1-c\right)}{\pi^{2}N}.
\end{equation}
The pure-GNY result agrees with Table 1 of Refs.~\cite{Iliesiu_2016,Iliesiu_2018}. 
The inverse correlation length exponent, $\nu^{-1}=d-\Delta_{\phi^{2}}$, is then
\begin{equation}
\nu^{-1}=1+\frac{16\left(c-1\right)}{\pi^{2}N}.
\end{equation}
Note that, for the chiral Ising QED$_{3}$ GNY model studied in this paper, and also for pure GNY, $c$ is neither zero nor unity, and thus $\eta_{\phi}$ and $\nu^{-1}$ have a non-zero correction at $\mathcal{O}(1/N)$.
However, a theory where the large-$N$ coefficient for $\eta_{\phi}$ does vanish~\cite{GraceyCHGNY_2018} at $\mathcal{O}(1/N)$
is the chiral Heisenberg GNY model, where the scalar field is promoted to a vector multiplet.  

For the cases of GNY and QED$_{3}$ GNY, $\nu^{-1}$ becomes
\begin{align}
\label{eq:nuInv}
\nu^{-1} = \left\{\begin{array}{lll}
 1-16/(3\pi^{2}N), & \text{GNY}, \\
 1-48/(\pi^{2}N), & \text{QED}_{3}\text{-GNY\ (no AL)},  \\ 
 1+80/(\pi^{2}N), & \text{QED}_{3}\text{-GNY}.\end{array}\right.
\end{align}
Again, the pure-GNY result agrees with the literature~\cite{Hands_1991,ZinnJustin_1991}, and the QED$_{3}$ GNY result appearing in the second line, which has ignored the QED$_{3}$ AL diagrams, 
agrees with the literature~\cite{GraceyQED3CS_1993,Ihrig_Janssen_2018}.
Our new result is the third line in Eq.~(\ref{eq:nuInv}): the inverse correlation length exponent correct to $\mathcal{O}(1/N)$ in fixed $d=3$, where we have incorporated the QED$_{3}$ AL diagrams. 
The importance of the Aslamazov-Larkin contribution is highlighted in the fact that it now changes the sign of this critical exponent. 
Importantly, this contribution now ensures that, to $\mathcal{O}(1/N)$, $\nu^{-1}\geq0$ for all $N$, in agreement with the unitarity requirement discussed in Sec.~\ref{sec:ScalingDim}.
However, the other unitarity requirement $\nu^{-1}\leq1+d/2=5/2$, in $d=3$ dimensions, sets a lower bound on $N$ of $N\geq6$, which does not include the value $N=2$ necessary to corroborate the conjectured duality in Eq.~(\ref{eq:And1}).

\section{Conclusion}
\label{sec:Conclusion}

\begin{figure}[t]
\centering\includegraphics[width=6cm,height=2.5cm,clip]{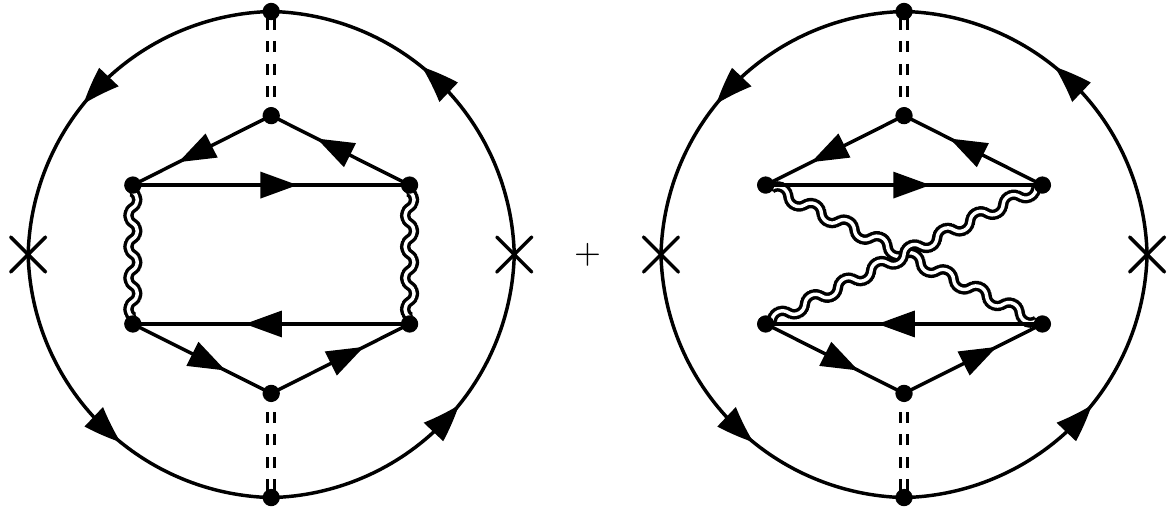}
\caption{Example of an $\mathcal{O}(1/N^2)$ correction to the fermion four-point function from QED$_3$ AL subdiagrams.}
\label{fig:MT_ALSubdiagram_Correction}
\end{figure} 

In this paper we have performed a $1/N$-expansion analysis of QED$_{3}$ GNY in $d=3$ spacetime dimensions, defined as in the duality conjecture of Ref.~\cite{Wang_2017} with $N$ two-component Dirac fermions and a Yukawa coupling preserving the 
SU$(N)$ flavor symmetry of pure QED$_3$. Our result for the fermion adjoint bilinear scaling dimension agrees at $\mathcal{O}(1/N)$ with the recent large-$N$ analysis of Ref.~\cite{GraceyQED3GNY_2018}, 
however, for the fermion singlet scaling dimension we showed that there is an additional contribution arising from QED$_{3}$ AL diagrams, which arises solely because of the three-dimensional gamma matrix tensor structure. 
These diagrams not only give crucial contributions to the scalar anomalous dimension and the correlation length exponent at $\mathcal{O}(1/N)$, but might also give new contributions to the adjoint bilinear scaling dimension at higher orders in $1/N$. 
For an example of such diagrams, see Fig.~\ref{fig:MT_ALSubdiagram_Correction}. Extrapolation of the results to $N=2$ using $\Pade$ and Borel-$\Pade$ approximants gave values for the scalar anomalous dimension and adjoint fermion bilinear scaling dimension in reasonable agreement with numerical studies of critical exponents at the N\'eel-to-VBS transition, in support of the duality conjecture. 
The explicit $\mathcal{O}(1/N)$ value of the inverse correlation length exponent in the QED$_3$ GNY model at $N=2$ was not within unitarity bounds and resummation methods were inapplicable. Our work nonetheless shows that it is vital to incorporate diagrams specific to $d=3$ when considering critical exponents in QED-like theories with two-component Dirac fermions, and their inclusion at higher orders in powers of $1/N$ may facilitate further understanding of the proposed $d=3$ dualities involving such theories.

{\it Note added.}---After this work was completed, we became aware of recent work~\cite{Benvenuti2018} which partially overlaps with ours.

\acknowledgments

We thank L. Di Pietro, S. Giombi, J. A. Gracey, L. Janssen, A. Penin, and C. Xu for helpful discussions. RB was supported by the Theoretical Physics Institute at the University of Alberta. AR was supported by NSERC and the Government of Alberta. JM was supported by NSERC grant \#RGPIN-2014-4608, the CRC Program, CIFAR, and the University of Alberta.

\appendix
\numberwithin{equation}{section}
\numberwithin{figure}{section}
\section{Fermion four-point correlation function}
This excursus gives an outline of the evaluation of the diagrams appearing in Sec.~\ref{sec:Fermion_CorrFunc}. To perform these calculations a list of core integrals that are continually referred to will be necessary; these are summarized below:
\begin{align}
\label{eq:Int_identity1}&\int\frac{d^{3}k}{\left(2\pi\right)^{3}}\frac{1}{k^{2}\left(k+p\right)^{2}}  = \frac{1}{8\left|p\right|},\\
\label{eq:Int_identity2}&\int\frac{d^{3}k}{\left(2\pi\right)^{3}}\frac{1}{k^{2}\left(k+p\right)^{2}\left(k+q\right)^{2}}  = \frac{1}{8\left|p\right|\left|q\right|\left|p-q\right|},\\
\label{eq:Int_identity3}&\int\frac{d^{3}k}{\left(2\pi\right)^{3}}\frac{1}{k^{2}\left(k+p\right)^{2}\left(k+q\right)^{2}\left(k+p+q\right)^{2}}\no\\
&\quad = \frac{1}{8\left|p\right|\left|q\right|p\cdot q}\left(\frac{1}{\left|p-q\right|}-\frac{1}{\left|p+q\right|}\right).
\end{align}
Feynman parameters can be used to derive Eqs.~(\ref{eq:Int_identity1}) and (\ref{eq:Int_identity2}), and Eq.~(\ref{eq:Int_identity3}) follows by using partial fractions to decompose the integrand and then using Eq.~(\ref{eq:Int_identity2}). Two other results are
\begin{align}
\label{eq:Int_identity4}&\int\frac{d^{3}k}{\left(2\pi\right)^{3}}\frac{2k\cdot q}{k^{4}\left(k+q\right)^{2}} = -\frac{1}{8\left|q\right|},\\
\label{eq:Int_identity5}&\int\frac{d^{3}k}{\left(2\pi\right)^{3}}\frac{2k\cdot q}{k^{4}\left(k+p\right)^{2}\left(k+q\right)^{2}} = -\frac{p^2+p\cdot q}{8p^2\left|p\right|\left|q\right|\left|p-q\right|}.
\end{align}
These can be derived by using differentiation under the integral sign, integration by parts, and then applying Eqs.~(\ref{eq:Int_identity1}) and (\ref{eq:Int_identity2}). 
Since the derivation of the QED$_{3}$ results presented in Eqs.~(\ref{eq:QED3_ABC},\ref{eq:QED3_DE}) is well documented~\cite{Rantner_Wen_2002,Chester_Pufu_2016}, 
we omit further discussion and defer to these articles, whose results we have independently confirmed.

\subsection{GNY diagrams}
\label{app:GNY}

Let $p$ denote the external momentum, assumed to flow from right to left. The diagrams in Figs.~\ref{fig:GNY_Diagrams}(a,c) are then
\begin{widetext}
\begin{align}
\mathcal{M}(p)_{6\A} &= -\frac{1}{N}\int\frac{d^{3}q}{\left(2\pi\right)^{3}}\frac{8}{\left|q\right|}\int\frac{d^{3}k}{\left(2\pi\right)^{3}}\frac{\mathrm{tr}\left[\slashed{k}\left(\slashed{k}+\slashed{q}\right)\left(\slashed{k}+\slashed{p}+\slashed{q}\right)\left(\slashed{k}+\slashed{p}\right)\right]}{k^{2}\left(k+q\right)^{2}\left(k+p+q\right)^{2}\left(k+p\right)^{2}},\\
\mathcal{M}(p)_{6\C} &= -\frac{1}{N}\int\frac{d^{3}q}{\left(2\pi\right)^{3}}\frac{8}{\left|q\right|}\int\frac{d^{3}k}{\left(2\pi\right)^{3}}\frac{\mathrm{tr}\left[\slashed{k}\left(\slashed{k}+\slashed{q}\right)\slashed{k}\left(\slashed{k}+\slashed{p}\right)\right]}{k^{4}\left(k+q\right)^{2}\left(k+p\right)^{2}}.
\end{align}
Note that, $\mathcal{M}(p)_{6\B}=\mathcal{M}(-p)_{6\C}=\mathcal{M}(p)_{6\C}$. 
To perform the trace algebra in the numerators of these two expressions, we use the result~\cite{Peskin_1995}:
\label{eq:Gamma4_Trace}
$\mathop{\mathrm{tr}}\gamma_{\alpha}\gamma_{\beta}\gamma_{\rho}\gamma_{\sigma}=2\left(\delta_{\alpha\beta}\delta_{\rho\sigma}+\delta_{\alpha\sigma}\delta_{\beta\rho}-\delta_{\alpha\rho}\delta_{\beta\sigma}\right).$
After taking the trace, the above expressions are then
\begin{align}
\mathcal{M}(p)_{6\A} &= -\frac{1}{N}\int\frac{d^{3}q}{\left(2\pi\right)^{3}}\frac{8}{\left|q\right|}\int\frac{d^{3}k}{\left(2\pi\right)^{3}}\frac{2k^{2}\left(p\cdot q+k\cdot p+\frac{1}{2}p^{2}\right)+2\left(k+q\right)^{2}\left(k\cdot p+k^{2}+\frac{1}{2}p^{2}\right)-q^{2}p^{2}}{k^{2}\left(k+q\right)^{2}\left(k+p+q\right)^{2}\left(k+p\right)^{2}},\\
\mathcal{M}(p)_{6\C} &= -\frac{1}{N}\int\frac{d^{3}q}{\left(2\pi\right)^{3}}\frac{8}{\left|q\right|}\int\frac{d^{3}k}{\left(2\pi\right)^{3}}\frac{2\left(k^{4}+k^{2}k\cdot q+k^{2}k\cdot p+2k\cdot pk\cdot q-k^{2}p\cdot q\right)}{k^{4}\left(k+q\right)^{2}\left(k+p\right)^{2}}.
\end{align}
\end{widetext}

To compute these integrals the method implemented is the same as that in concomitant literature~\cite{WeiChen_1992,WeiChen_1993,Rantner_Wen_2002}. The integration over $k$ is convergent, and so this is performed first. 
The crucial step is to decompose all the dot products involving the integration variable $k$ into a sum of squares, as done in the Passarino-Veltman decomposition~\cite{Passarino_1979,Denner_2006,SmirnovBook}; for instance, $2k\cdot p = (k+p)^2-k^2-p^2$. 
Performing this reduction on all dot products involving $k$ then results in a collection of integrals that now have only quadratic (or quartic) functions of the momenta $k$, which necessarily facilitates the use of the general identities listed earlier. 
The remaining $k$-integrals can then all be computed using the integration identities in Eqs.~(\ref{eq:Int_identity1}-\ref{eq:Int_identity5}).
After this procedure, one obtains
\begin{align}
\mathcal{M}(p)_{6\A} &= -\frac{2}{N}\int\frac{d^{3}q}{\left(2\pi\right)^{3}}\frac{1}{\left|q\right|\left|p+q\right|}\left(1+\frac{\left|p\right|\left|q\right|}{p\cdot q}\right),\\
\mathcal{M}(p)_{6\C} &= -\frac{1}{N}\int\frac{d^{3}q}{\left(2\pi\right)^{3}}\frac{1}{\left|q\right|\left|p+q\right|}\left(1+\frac{p\cdot q}{\left|p\right|\left|q\right|}\right).
\end{align}
The total contribution from the GNY diagrams in Figs.~\ref{fig:GNY_Diagrams}(a-c) is $\mathcal{M}(p)_{6\A+6\B+6\C}=\mathcal{M}(p)_{6\A}+2\mathcal{M}(p)_{6\C}$: 
\begin{align}
\mathcal{M}(p)_{6\A+6\B+6\C} 
&= -\frac{2}{N}\int\frac{d^{3}q}{\left(2\pi\right)^{3}}\frac{1}{\left|q\right|\left|p+q\right|}\nonumber\\
&\quad\times\biggl(2+\frac{p\cdot q}{\left|p\right|\left|q\right|}+\frac{\left|p\right|\left|q\right|}{p\cdot q}\biggr).
\end{align}
The presence of the dot product term $p\cdot q$ in the denominator of the third term occludes implementing dimensional regularization as a regularization technique for the $q$-integration~\cite{WeiChen_1992,WeiChen_1993,Rantner_Wen_2002}.
However, since the integration is over only a single variable $q$, spherical-polar coordinates can be used, along with a UV cutoff $\Lambda$ for the radial integration. 
The use of this UV cutoff may possibly lead to quadratically and linearly divergent terms in $\Lambda$, which are of no physical importance, and merely arise due to the choice of a regulator that breaks gauge invariance~\cite{WeiChen_1992,WeiChen_1993,Rantner_Wen_2002}. 
The first two terms give: 
\begin{align}
I_{1}  = -\frac{\left|p\right|}{2\pi^{2}N}\left[\frac{16}{9}+4\left(\frac{\Lambda}{\left|p\right|}-1\right)-\frac{1}{3}\ln\left(\frac{\Lambda^{2}}{p^{2}}\right)\right].
\end{align}
The remaining integral can be computed to give
\begin{align}
I_{2} &= -\frac{\left|p\right|}{4\pi^{2}N}\biggl[\int_{0}^{1}\frac{dy}{\sqrt{1+y}}\ln\left(\frac{\sqrt{1+y}-\sqrt{y}}{\sqrt{1+y}+\sqrt{y}}\right)\nonumber\\
&\quad+\int_{1}^{\Lambda^{2}/p^{2}}\frac{dy}{\sqrt{1+y}}\ln\left(\frac{\sqrt{1+y}-1}{\sqrt{1+y}+1}\right)\biggr],\nonumber
 \end{align}
 \begin{align}&= -\frac{\left|p\right|}{4\pi^{2}N}\biggl\{4\left(1+\sqrt{2}\ln\left(\sqrt{2}-1\right)\right)\nonumber\\
 &\quad +2\sqrt{1+\Lambda^{2}/p^{2}}\ln\left(\frac{\sqrt{1+\Lambda^{2}/p^{2}}-1}{\sqrt{1+\Lambda^{2}/p^{2}}+1}\right)\nonumber\\
 &\quad-4\sqrt{2}\ln\left(\sqrt{2}-1\right)-2\ln\left(\frac{\Lambda^{2}}{p^{2}}\right)\biggr\}.
\end{align}
The first expression arises from the integration over the domain $[0,1]$ whereas the remaining expressions arise from the integration over the domain $[1,\Lambda^2/p^2]$. In the limit that $\Lambda/\left|p\right|\gg1$, the above result has the closed form:
\begin{equation}
I_{2} = \frac{\left|p\right|}{2\pi^{2}N}\ln\left(\frac{\Lambda^{2}}{p^{2}}\right).
\end{equation}
Combining the above two results together (and ignoring the linearly divergent part) gives Eq.~(\ref{eq:GNY_ABC}) of the main text
\begin{equation}
\mathcal{M}(p)_{6\A+6\B+6\C}=\frac{2|p|}{3\pi^2N}\ln\left(\frac{\Lambda^2}{p^2}\right).
\end{equation}

The remaining pure-GNY diagrams to compute are those shown in Figs.~\ref{fig:GNY_Diagrams}(d-e). In what follows it will be shown that the triangle vertex appearing in these diagrams vanishes. The diagram in Fig.~\ref{fig:GNY_Diagrams}(d) is given by
\begin{equation}
\mathcal{M}(p)_{6\D}=\int\frac{d^{3}q}{\left(2\pi\right)^{3}}V_{1}\left(p,q\right)D_{\phi}\left(q\right)V_{2}\left(p,q\right)D_{\phi}\left(p-q\right).
\end{equation}
Using the identity $\mathop{\mathrm{tr}}\gamma_{\alpha}\gamma_{\beta}\gamma_{\sigma}=2i\epsilon_{\alpha\beta\sigma}$, the left-most triangle vertex is given by
\begin{align}
V_{1}\left(p,q\right) &= \frac{-i}{\sqrt{N}}\int\frac{d^{3}k}{\left(2\pi\right)^{3}}\frac{\mathrm{tr}\left[\left(\slashed{k}+\slashed{p}\right)\left(\slashed{k}+\slashed{q}\right)\slashed{k}\right]}{k^{2}\left(k+p\right)^{2}\left(k+q\right)^{2}},\nonumber\\
& = \frac{2}{\sqrt{N}}\int\frac{d^{3}k}{\left(2\pi\right)^{3}}\frac{\epsilon_{\alpha\beta\sigma}p_{\alpha}q_{\beta}k_{\sigma}}{k^{2}\left(k+p\right)^{2}\left(k+q\right)^{2}}.
\end{align}
The denominator can be rewritten using the Feynman parameters $x$ and $y$:
\begin{equation}
\frac{1}{k^{2}\left(k+p\right)^{2}\left(k+q\right)^{2}}=\int_{0}^{1}dx\int_{0}^{1-x}dy\frac{2}{\left(L^{2}+\Delta\right)^{3}},
\end{equation}
where $L=k+xp+yq, \Delta=xp^{2}+yq^{2}-\left(xp+yq\right)^{2}$. 
Thus, the triangle vertex now becomes: 
\begin{align}
V_{1}\left(p,q\right) &= \epsilon_{\alpha\beta\sigma}p_{\alpha}q_{\beta}\frac{4}{\sqrt{N}}\int_{0}^{1}dx\int_{0}^{1-x}dy\nonumber\\
&\quad\times\int\frac{d^{d}L}{\left(2\pi\right)^{d}}\frac{\left(L-xp-yq\right)_{\sigma}}{\left(L^{2}+\Delta\right)^{3}}=0.
\end{align}
Since the triangle vertex is zero, both pure-GNY AL diagrams vanish; this confirms the result in Eq.~(\ref{eq:GNY_DE}) of the main text:
\begin{equation}
\mathcal{M}(p)_{6\D+6\E}=0.
\end{equation}

\subsection{QED$_{3}$ GNY diagrams}
\label{app:QED3_GNY}

The only new diagrams that need to be considered for QED$_{3}$ GNY are the gauge-scalar AL diagrams shown in Figs.~\ref{fig:QED3GNY_Diagrams}(a-d); the first such diagram is
\begin{equation}
\mathcal{M}(p)_{7\A}=\int\frac{d^{3}q}{\left(2\pi\right)^{3}}V_{1}^{\mu}\left(p,q\right)\Pi_{\mu\nu}\left(q\right)V_{2}^{\nu}\left(p,q\right)D_{\phi}\left(p-q\right).
\end{equation}
The left-most triangle vertex is given by
\begin{equation}
V_{1}^{\mu}\left(p,q\right)=\frac{1}{\sqrt{N}}\int\frac{d^{3}k}{\left(2\pi\right)^{3}}\frac{\mathrm{tr}\left[\left(\slashed{k}+\slashed{p}\right)
\left(\slashed{k}+\slashed{q}\right)\gamma_{\mu}\slashed{k}\right]}{k^{2}\left(k+p\right)^{2}\left(k+q\right)^{2}}.
\end{equation}
By permuting the gamma matrices, and using the identity $\mathop{\mathrm{tr}}\gamma_{\alpha_{1}}\cdots\gamma_{\alpha_{n}}=\left(-1\right)^{n}\mathop{\mathrm{tr}}\gamma_{\alpha_{n}}\cdots\gamma_{\alpha_{1}}$,
it follows that the right-most triangle vertex obeys $V_{2}^{\nu}\left(p,q\right)=V_{1}^{\nu}\left(p,q\right)$. This is shown below:
\begin{align}
\label{eq:V2vertex}
V_{2}^{\nu}\left(p,q\right)&=\frac{1}{\sqrt{N}}\int\frac{d^{3}l}{\left(2\pi\right)^{3}}\frac{\mathrm{tr}\left[\slashed{l}\gamma_{\nu}
\left(\slashed{l}+\slashed{q}\right)\left(\slashed{l}+\slashed{p}\right)\right]}{l^2\left(l+q\right)^{2}\left(l+p\right)^{2}},\no\\
&=\frac{1}{\sqrt{N}}\int\frac{d^{3}l}{\left(2\pi\right)^{3}}\frac{\mathrm{tr}\left[\left(\slashed{l}+\slashed{p}\right)\left(\slashed{l}+\slashed{q}\right)\gamma_{\nu}\slashed{l}\right]}{l^2\left(l+p\right)^{2}\left(l+q\right)^{2}},\no\\
&=V_{1}^{\nu}\left(p,q\right).
\end{align}

The trace algebra is readily computed, and thus the triangle vertex becomes
\begin{align}
V_{1}^{\mu}\left(p,q\right) &= \frac{2}{\sqrt{N}}\int\frac{d^{3}k}{\left(2\pi\right)^{3}}\frac{1}{k^{2}\left(k+p\right)^{2}\left(k+q\right)^{2}}&\no\\
&\quad \times[k^{2}\left(k_{\mu}-p_{\mu}+q_{\mu}\right)+\left(2k\cdot p+p\cdot q\right)k_{\mu}\no\\
&\quad +k\cdot pq_{\mu}-k\cdot qp_{\mu}].
\end{align}

To compute these integrals, the compendium of results given below are required:
\begin{align}
&\int\frac{d^{3}k}{\left(2\pi\right)^{3}}\frac{k_{\mu}}{k^{2}\left(k+p\right)^{2}\left(k+q\right)^{2}}\no\\
&\quad= \frac{\alpha_{0}p_{\mu}+\beta_{0}q_{\mu}}{p^{2}q^{2}-\left(p\cdot q\right)^{2}}\equiv I_{0}^{\mu}\left(p,q\right), \\
&\alpha_{0} = \frac{q^{2}}{16}\left(\frac{1}{\left|q\right|}-\frac{1}{\left|p-q\right|}-\frac{\left|p\right|}{\left|q\right|\left|p-q\right|}\right)\no\\
&\quad\quad-\frac{p\cdot q}{16}\left(\frac{1}{\left|p\right|}-\frac{1}{\left|p-q\right|}-\frac{\left|q\right|}{\left|p\right|\left|p-q\right|}\right), \\
&\beta_{0} = \frac{p^{2}}{16}\left(\frac{1}{\left|p\right|}-\frac{1}{\left|p-q\right|}-\frac{\left|q\right|}{\left|p\right|\left|p-q\right|}\right)\no\\
&\quad\quad -\frac{p\cdot q}{16}\left(\frac{1}{\left|q\right|}-\frac{1}{\left|p-q\right|}-\frac{\left|p\right|}{\left|q\right|\left|p-q\right|}\right),\\
&\int\frac{d^{3}k}{\left(2\pi\right)^{3}}\frac{k\cdot pk_{\mu}}{k^{2}\left(k+p\right)^{2}\left(k+q\right)^{2}}\no\\
&\quad= \frac{1}{32}\left(-\frac{q_{\mu}}{\left|q\right|}+\frac{p_{\mu}+q_{\mu}}{\left|p-q\right|}-16p^{2}I_{0}^{\mu}\left(p,q\right)\right), \\
&\int\frac{d^{3}k}{\left(2\pi\right)^{3}}\frac{k_{\mu}}{\left(k+p\right)^{2}\left(k+q\right)^{2}} = -\frac{1}{16}\frac{p_{\mu}+q_{\mu}}{\left|p-q\right|}.
\end{align}
These results can all be derived by using the Passarino-Veltman decomposition and applying the identities given in Eqs.~(\ref{eq:Int_identity1}-\ref{eq:Int_identity2}).
The integrals appearing in the triangle vertex are then readily calculated, and the final result is
\begin{align}
V_{1}^{\mu}\left(p,q\right) &= -\frac{1}{8\sqrt{N}}\biggl[\frac{p_{\mu}-q_{\mu}}{\left|p-q\right|}+16\left(p^{2}-p\cdot q\right)I_{0}^{\mu}\left(p,q\right)\no\\&\quad
+\frac{q_{\mu}\left|p\right|}{\left|q\right|\left|p-q\right|}+\frac{p_{\mu}}{\left|p\right|}\left(1-\frac{\left|q\right|}{\left|p-q\right|}\right)\biggr].
\end{align}
By gauge invariance $q_{\mu}\Pi_{\mu\nu}\left(q\right)=\Pi_{\mu\nu}\left(q\right)q_{\nu}=0$, thus all the terms in $V_{1}^{\mu}$ and $V_{1}^{\nu}$ proportional to $q_{\mu}$ or $q_{\nu}$ give zero contribution to the final answer. 
Thus, these terms can be dropped so that the triangle vertex simplifies  to:
\begin{align}
V_{1}^{\mu}\left(p,q\right) &= -\frac{p_{\mu}}{8\sqrt{N}}\biggl[\frac{1}{\left|p-q\right|}+\frac{1}{\left|p\right|}\left(1-\frac{\left|q\right|}{\left|p-q\right|}\right)\no\\
&\quad+16\alpha_{0}\frac{p^{2}-p\cdot q}{p^{2}q^{2}-\left(p\cdot q\right)^{2}}\biggr],\no\\
& \equiv \frac{p_{\mu}}{8\sqrt{N}}F\left(p,q\right).
\end{align}
Let $\theta$ denote the angle between the vectors $p$ and $q$; a useful identity is 
\begin{equation}
p_{\mu}\Pi_{\mu\nu}\left(q\right)p_{\nu}=\frac{16}{\left|q\right|}p_{\mu}\left(\delta_{\mu\nu}-\frac{q_{\mu}q_{\nu}}{q^{2}}\right)p_{\nu}=\frac{16}{\left|q\right|}p^{2}\sin^{2}\theta.
\end{equation}
Performing the $q$ integration, the gauge-scalar AL diagram is as presented in Eq.~(\ref{eq:QED3GNY_A}) of the main text:
\begin{align}
\mathcal{M}(p)_{7\A} &= \frac{2p^{2}}{N}\int\frac{d^{3}q}{\left(2\pi\right)^{3}}\frac{\sin^{2}\theta}{\left|p+q\right|\left|q\right|}F\left(p,-q\right)^{2},\nonumber\\
&= \frac{p^{2}}{2\pi^{2}N}\int_{0}^{\infty}dQQ\int_{-1}^{1}dx\frac{F\left(p,-q\right)^{2}(1-x^{2})}{\sqrt{p^{2}+Q^{2}+2\left|p\right|Qx}},\nonumber\\
&= \frac{2\left|p\right|}{3N}\left(1-\frac{9}{\pi^{2}}\right).
\end{align}

The diagram shown in Fig.~\ref{fig:QED3GNY_Diagrams}(c) is 
\begin{equation}
\mathcal{M}(p)_{7\C}=\int\frac{d^{3}q}{\left(2\pi\right)^{3}}V_{1}^{\mu}\left(p,q\right)\Pi_{\mu\nu}\left(q\right)V_{3}^{\nu}\left(p,q\right)D_{\phi}\left(p-q\right).
\end{equation}
In contrast to Fig.~\ref{fig:QED3GNY_Diagrams}(a), however, the right-most triangle vertex is now:
\begin{align}
V_{3}^{\nu}\left(p,q\right)&=\frac{1}{\sqrt{N}}\int\frac{d^{3}l}{\left(2\pi\right)^{3}}\frac{\mathrm{tr}\left[\left(\slashed{l}-\slashed{p}\right)\left(\slashed{l}-\slashed{q}\right)\gamma_{\nu}\slashed{l}\right]}{\left(l-p\right)^{2}\left(l-q\right)^{2}l^{2}},\no\\
&=-\frac{1}{\sqrt{N}}\int\frac{d^{3}l}{\left(2\pi\right)^{3}}\frac{\mathrm{tr}\left[\left(\slashed{l}+\slashed{p}\right)
\left(\slashed{l}+\slashed{q}\right)\gamma_{\nu}\slashed{l}\right]}{l^{2}\left(l+p\right)^{2}\left(l+q\right)^{2}},\no\\
&=-V_{1}^{\nu}\left(p,q\right).
\end{align}
Importantly, in comparison with Eq.~(\ref{eq:V2vertex}) there is a relative sign difference between the two vertices. Thus, the two diagrams are equal and opposite: 
\begin{equation}
\mathcal{M}(p)_{7\C}=-\mathcal{M}(p)_{7\A}.
\end{equation}
A similar calculation shows that $\text{7(b)} =- \text{7(d)}$. Therefore, the QED$_{3}$ GNY gauge-scalar AL diagrams sum to zero:
\begin{equation}
\mathcal{M}(p)_{7\A+7\B+7\C+7\D}=0.
\end{equation}
\section{Scalar correlation functions}
\label{app:BoxFunctions}

There are three types of fermion ``box'' functions in pure-GNY which contribute to the scalar four-point function $G_{\phi^{2}}$: two ``box'' (B), two ``bow-tie'' (BT), and two ``hourglass'' (HG) diagrams. 
The B and BT diagrams are equivalent, however, the same is not true for the B and HG diagrams. Importantly, the latter two have different logarithmically singular contributions. 
In Secs.~\ref{app:FermionBox}-\ref{app:FermionHourGlass} we present derivations of the logarithmically singular part of each diagram.

\subsection{Fermion box diagram }
\label{app:FermionBox}

The fermion box diagram $I_{\BO}$ appearing in Fig.~\ref{fig:Box_BowTie_HourGlass_Diagrams}(a) is
\begin{align}
I_{\BO}&=-\frac{8^{4}}{N}\int\frac{d^{3}k}{\left(2\pi\right)^{3}}\int\frac{d^{3}l}{\left(2\pi\right)^{3}}\int\frac{d^{3}q}{\left(2\pi\right)^{3}}
\frac{1}{\left|k\right|\left|k+p\right|}\frac{1}{\left|l\right|\left|l+p\right|}\no\\&\quad
\times\frac{\mathrm{tr}\left[\slashed{q}\left(\slashed{q}-\slashed{k}\right)\left(\slashed{q}+\slashed{p}\right)\left(\slashed{q}-\slashed{l}\right)\right]}{q^{2}\left(q-k\right)^{2}\left(q+p\right)^{2}\left(q-l\right)^{2}}.
\end{align}
The trace algebra in the numerator can be computed exactly, and the result is 
\begin{align}
\mathrm{Num} &= \mathrm{tr}\left[\slashed{q}\left(\slashed{q}-\slashed{k}\right)\left(\slashed{q}+\slashed{p}\right)\left(\slashed{q}-\slashed{l}\right)\right] ,\nonumber\\ 
 &= 2[q^{4}-q^{2}\left(l\cdot q+l\cdot p+k\cdot q+k\cdot p+l\cdot k-p\cdot q\right)\no\\
 &\quad+2l\cdot qk\cdot q+l\cdot pk\cdot q+k\cdot pl\cdot q-l\cdot kp\cdot q].
\end{align}
In the limit $\left|p\right|\ll\Lambda$, the numerator reduces to 
\begin{equation}
\mathrm{Num}\approx2\left[q^{4}-q^{2}\left(l\cdot q+k\cdot q+l\cdot k\right)+2l\cdot qk\cdot q\right].
\end{equation}

In order to take the limit $\left|p\right|\ll\Lambda$ in the integrand, we write 
\begin{equation}
\frac{1}{\left|k+p\right|\left|l+p\right|}=\frac{1}{\left|k\right|\left|l+p\right|}+\frac{1}{\left|k+p\right|\left|l\right|}-\frac{1}{\left|k\right|\left|l\right|}+\mathcal{O}\left(p^{2}\right).\label{eq:Denom_Exp}
\end{equation}
This approximation is correct to linear order in $\left|p\right|$. Since the integrand is symmetric in $k$ and $l$, the first two terms
give equal contributions, while the third term is independent of external momentum $p$, and thus can be ignored. 
Thus, the box diagram is
\begin{align}
I_{\BO}&=-\frac{4\times8^{4}}{N}\int\frac{d^{3}l}{\left(2\pi\right)^{3}}\frac{1}{\left|l\right|\left|l+p\right|}\int\frac{d^{3}q}{\left(2\pi\right)^{3}}\int\frac{d^{3}k}{\left(2\pi\right)^{3}}\frac{1}{k^{2}}
\no\\&\quad\times\frac{q^{4}-q^{2}\left(l\cdot q+k\cdot q+l\cdot k\right)+2l\cdot qk\cdot q}{q^{4}\left(q-k\right)^{2}\left(q-l\right)^{2}}.\label{eq:I_Box_1}
\end{align}
The $k$ and $q$ integrals of the five terms above can all be computed using the Passarino-Veltman decomposition and the identities in Eqs.~(\ref{eq:Int_identity1}-\ref{eq:Int_identity5}).
Inserting the result into Eq.~(\ref{eq:I_Box_1}) then produces 
\begin{align}
I_{\BO} &= -\frac{4\times8^{4}}{16\left(2\pi\right)^{2}N}\int\frac{d^{3}l}{\left(2\pi\right)^{3}}\frac{1}{\left|l\right|\left|l+p\right|}\ln\left(\frac{\Lambda^{2}}{l^{2}}\right),\nonumber \\
 &= \frac{4\times8^{2}}{\pi^{2}N}\frac{\left|p\right|}{\left(2\pi\right)^{2}}\ln\left(\frac{\Lambda^{2}}{p^{2}}\right).
\end{align}
Here we have retained only the logarithmically singular contribution. The scalar-scalar bubble is given in Eq.~(\ref{eq:Boson_Bubble}), and in terms of this the logarithmically singular part of the box diagram is
\begin{equation}
I_{\BO}=-G_{\phi^{2},9\A}\left(p\right)\frac{2}{\pi^{2}N}\ln\left(\frac{\Lambda^{2}}{p^{2}}\right).
\end{equation}

\subsection{Fermion bow-tie diagram }
\label{app:FermionBowTie}

The fermion bow-tie diagram $I_{\BT}$ appearing in Fig.~\ref{fig:Box_BowTie_HourGlass_Diagrams}(c) is
\begin{align}
I_{\BT}&=\frac{-8^4}{N}\int\frac{d^{3}k}{\left(2\pi\right)^{3}}\int\frac{d^{3}l}{\left(2\pi\right)^{3}}\int\frac{d^{3}q}{\left(2\pi\right)^{3}}\frac{1}{\left|k\right|\left|k+p\right|}\frac{1}{\left|l\right|\left|l+p\right|}
\no\\&\quad\times\frac{\mathrm{tr}\left[\slashed{q}\left(\slashed{q}-\slashed{k}\right)\left(\slashed{q}+\slashed{p}\right)\left(\slashed{q}+\slashed{l}+\slashed{p}\right)\right]}{q^{2}\left(q-k\right)^{2}\left(q+p\right)^{2}\left(q+l+p\right)^{2}}.
\end{align}
Shifting $l\rightarrow-l-p$ produces the expression for the fermion box diagram, thus, using the previously obtained results:
\begin{equation}
I_{\BT}=-G_{\phi^{2},9\A}\left(p\right)\frac{2}{\pi^{2}N}\ln\left(\frac{\Lambda^{2}}{p^{2}}\right).
\end{equation}

\subsection{Fermion hourglass diagram}
\label{app:FermionHourGlass}

The fermion hourglass diagram appearing in Fig.~\ref{fig:Box_BowTie_HourGlass_Diagrams}(e) is
\begin{align}
I_{\HG}&=\frac{-8^{4}}{N}\int\frac{d^{3}k}{\left(2\pi\right)^{3}}\int\frac{d^{3}l}{\left(2\pi\right)^{3}}\int\frac{d^{3}q}{\left(2\pi\right)^{3}}\frac{1}{\left|k\right|\left|k+p\right|}\frac{1}{\left|l\right|\left|l+p\right|}
\no\\&\quad\times\frac{\mathrm{tr}\left[\left(\slashed{q}-\slashed{k}-\slashed{l}\right)\left(\slashed{q}-\slashed{k}\right)\left(\slashed{q}+\slashed{p}\right)\left(\slashed{q}-\slashed{l}\right)\right]}{\left(q-k-l\right)^{2}\left(q-k\right)^{2}\left(q+p\right)^{2}\left(q-l\right)^{2}}.
\end{align}
In the limit $\left|p\right|\ll\Lambda$, the numerator becomes

\begin{align}
\mathrm{Num} &\approx \mathrm{tr}\left[\left(\slashed{q}-\slashed{k}-\slashed{l}\right)\left(\slashed{q}-\slashed{k}\right)\slashed{q}\left(\slashed{q}-\slashed{l}\right)\right],\nonumber \\
 &= 2[q^{4}-q^{2}\left(2k\cdot q+2l\cdot q-l^{2}-k^{2}-k\cdot l\right)
 \no\\&\quad-l^{2}k\cdot q-k^{2}l\cdot q+2k\cdot ql\cdot q].
\end{align}
The integrand is again expanded as in Eq.~(\ref{eq:Denom_Exp}) and due to the symmetry of the integrand, the hourglass diagram reduces to 
\begin{widetext}
\begin{align}
I_{\HG} &= -\frac{4\times8^{4}}{N}\int\frac{d^{3}l}{\left(2\pi\right)^{3}}\frac{1}{\left|l\right|\left|l-p\right|}\int\frac{d^{3}q}{\left(2\pi\right)^{3}}\int\frac{d^{3}k}{\left(2\pi\right)^{3}}\frac{1}{k^{2}}\nonumber \\
 & \quad\times \biggl[\frac{q^{4}+q^{2}\left(2k\cdot q+2l\cdot q+l^{2}+k^{2}+k\cdot l\right)+l^{2}k\cdot q+k^{2}l\cdot q+2k\cdot ql\cdot q}{q^{2}\left(q+k\right)^{2}\left(q+k+l\right)^{2}\left(q+l\right)^{2}}\biggr].
\end{align}
\end{widetext}

The nine integrals appearing here can be calculated and the ensuing result for the hourglass diagram is then
\begin{align}
I_{\HG} &= -\frac{4\times8^{4}}{8\left(2\pi\right)^{2}N}\int\frac{d^{3}l}{\left(2\pi\right)^{3}}\frac{1}{\left|l\right|\left|l+p\right|}\ln\left(\frac{\Lambda^{2}}{l^{2}}\right),\nonumber\\
&= -G_{\phi^{2},9\A}\left(p\right)\frac{4}{\pi^{2}N}\ln\left(\frac{\Lambda^{2}}{p^{2}}\right).
\end{align}
The hourglass diagram is twice the box diagram.
In total there are two box diagrams, two bow-tie diagrams, and two hourglass diagrams, as indicated in Figs.~\ref{fig:Box_BowTie_HourGlass_Diagrams}(a-f). Combining the previously obtained results then gives the total fermion box function as in Eq.~(\ref{eq:Fermion_Box}) of the main text:
\begin{equation}
G_{\phi^{2},\text{box}}\left(p\right)=-G_{\phi^{2},9\A}\left(p\right)\frac{16}{\pi^{2}N}\ln\left(\frac{\Lambda^{2}}{p^{2}}\right).
\end{equation}

\section{Pad\'e and Borel-Pad\'e resummation}
\label{sec:PBP}

In general, the $1/N$-expansion series, just like the $\epsilon$-expansion series, is an asymptotic series with zero radius of convergence~\cite{KleinertBook}. 
Thus, from a mathematical standpoint, resummation procedures, which are consistent ways of assigning a finite value to a formally divergent series, are necessary to extract physical results. The two most standard techniques are Pad\'e and Borel-Pad\'e resummation~\cite{KleinertBook}. For instance, those two techniques were recently employed to obtain small-$N$ estimates of critical exponents from $\mathcal{O}(1/N)$ and $\mathcal{O}(1/N^2)$ expressions for the chiral Ising- GNY and QED$_3$ GNY models (without AL diagrams)~\cite{GraceyQED3GNY_2018}. These methods will be utilized here as well to obtain estimates for the critical exponents and scaling dimensions for $N=2$.

For a given loop order $L$, the (one-sided) Pad\'e approximants are defined by  
\begin{equation}
[m/n](N)=\frac{\sum_{i=0}^m a_i/N^i}{1+\sum_{j=1}^nb_{j}/N^{j}},
\end{equation}
where $m$ and $n$ are two positive integers obeying $m+n=L$. The coefficients $a_i$ and $b_j$ are determined such that expanding the above function in powers of $1/N$ to $\mathcal{O}(1/N^{L})$ reproduces the $1/N$-expansion results. 
The Borel-Pad\'e transform is defined as follows. For an expansion given by $\Delta(N)=\sum_{k=0}^{\infty}\Delta_{k}/N^k$, the Borel sum is defined as $B_{\Delta}(N)=\sum_{k=0}^{\infty}\Delta_{k}/(N^{k}k!)$. The Borel-Pad\'e transform is then:
\begin{equation}
\Delta(N)=\int_{0}^{\infty}dt\ e^{-t}B_{\Delta}\left(t/N\right).
\end{equation}
The $1/N$-expansion coefficients have been computed here only to first order, and therefore in the above expression $B_{\Delta}$ is replaced by the [0/1] Pad\'e approximant. 

Using the expressions above for $N=2$, we obtain the critical exponents and scaling dimensions reported in Table~\ref{tab:Results} to four significant digits.

\bibliography{QED3GNY_LargeN_3d}
\end{document}